\documentclass[12pt]{article}
\usepackage{a4wide,amssymb,cite}
\pdfoutput=1
\usepackage{amsmath}
\usepackage{a4wide,amssymb,cite,graphicx}
\usepackage{epsfig}
\usepackage[usenames,dvipsnames]{color}
\usepackage{slashed}
\newcommand{\be}{\begin{equation}}
\newcommand{\ee}{\end{equation}}
\newcommand{\bea}{\begin{eqnarray}}
\newcommand{\eea}{\end{eqnarray}}

\def\circa#1{\,\raise.3ex\hbox{$#1$\kern-.75em\lower1ex\hbox{$\sim$}}\,}

\begin{document}

\begin{titlepage}
%
%


%

\begin{centering}
\vspace{1cm}
{\Large {\bf Cosmic abundances of SIMP dark matter}} \\

\vspace{1.5cm}

{\bf Soo-Min Choi$^\dagger$,  Hyun Min Lee$^\dagger$ and Min-Seok Seo$^\ddagger$}

\vspace{.2cm}

\vspace{.5cm}

{\it $^\dagger$Department of Physics, Chung-Ang University, Seoul 06974, Korea.} \\
{\it $^\ddagger$Center for Theoretical Physics of the Universe,
Institute for Basic Science, 34051 Daejeon, Korea.}

\end{centering}
\vspace{2cm}

\begin{abstract}
\noindent
Thermal production of light dark matter with sub-GeV scale mass can be attributed to $3\rightarrow 2$ self-annihilation processes. 
We consider the thermal average for annihilation cross sections of dark matter at $3\rightarrow 2$  and general higher-order interactions. A correct thermal average for initial dark matter particles is important, in particular, for annihilation cross sections with overall velocity dependence and/or resonance poles.
We apply our general results to benchmark models for SIMP dark matter and discuss the effects of the resonance pole in determining the relic density.  

\end{abstract}

\vspace{3cm}

\end{titlepage}

\section{Introduction}

Thermal dark matter, that was once in chemical equilibrium and decoupled from thermal plasma in the Universe, has been one of the plausible candidates for dark matter, in particular, under the name of Weakly Interacting Massive Particles (WIMP). 
Chemical equilibrium of dark matter usually requires a standard $2\rightarrow 2$ annihilation of dark matter into the SM particles, so it has provided an interesting interplay between the relic density, direct and indirect detection of dark matter at terrestrial and satellite experiments. Recently, a new mechanism for freezing out the density of dark matter from the $3\rightarrow 2$ annihilation process, coined the Strongly Interacting Massive Particles (SIMP) \cite{Hochberg1}, has recently drawn special attention, due to the fact that there is no need of a large coupling between dark matter and the SM particles in this case.

Dark matter in the early Universe has once had a Maxwell-Boltzmann velocity distribution in the non-relativistic limit for which the DM annihilates only. Thus, there is a need of making a thermal average for the annihilation cross section of dark matter in order to incorporate it in the Boltzmann equation for the DM relic density. In particular, when the annihilation cross section depends strongly on the DM velocity, for instance, due to dominance of higher partial waves or resonance poles. 
In the case of WIMP dark matter that is based on the $2\rightarrow 2$ annihilation, it is enough to do the thermal average for the velocity of a single DM particle or the relative velocity in the center of mass frame. On the other hand, in the case of SIMP dark matter that is based on the $3\rightarrow 2$ annihilation, we need to do the thermal averages for two (relative) velocities of dark matter in the initial states. Given that the velocity dependence of the $3\rightarrow 2$ annihilation depends on the properties of dark matter \cite{Hochberg2,z3global,simp2,z3simp,vsimp,z5simp,vsimp2} and the existence of resonance poles \cite{z5simp}, it is worthwhile to make a systematic study of the thermal averages for $3\rightarrow 2$ and higher-order annihilation processes in general.

In this article, we present a general discussion on the thermal average of the $3\rightarrow 2$ annihilation cross section in the perturbative regime where the velocity expansion is valid and near the resonance pole that mediates between three particles in the initial state and two particles in the final state. We discuss the effects of the resonance pole on the thermal-averaged cross section as well as the relic density and compare the results to the WIMP case.  Representative examples for SIMP dark matter, such as models with $Z_n$ discrete symmetries and dark mesons, are discussed in light of the thermal average of the $3\rightarrow 2$ annihilation cross section without or with a resonance pole.

The paper is organized as follows. 
We begin with a review on the thermal average of the $2\rightarrow 2$ annihilation cross section and then discuss a counterpart of the $3\rightarrow 2$ annihilation cross section without or with a resonance.  Next we incorporate the thermal-averaged cross sections in the Boltzmann equations for WIMP and SIMP cases and apply our general results for known models for SIMP dark matter. 
We continue to generalize our discussion to the $3\rightarrow 2$ coannihilation between particles with different masses and higher-order annihilation processes. 
Finally, conclusions are drawn.

\section{Thermal average for $2\rightarrow 2$ DM annihilations}

To warm up and compare to our later discussion on $3\rightarrow 2$ processes, we first give a review on the thermal average of the standard $2\rightarrow 2$ annihilation cross section without or with a resonance. 
Assuming that two DM particles in the initial states have the same masses, $m_1=m_2\equiv m_{\rm DM}$, the thermal averaged $2\rightarrow 2$ cross section is given by
\bea
\langle \sigma v\rangle= \frac{\int d^3 v_1 d^3 v_2 \, \delta^3({\vec v}_1+{\vec v}_2)(\sigma v)\, e^{-\frac{1}{2}x(v^2_1+v^2_2)}}{\int d^3 v_1 d^3 v_2 \, \delta^3({\vec v}_1+{\vec v}_2) \,e^{-\frac{1}{2}x(v^2_1+v^2_2)}},
\eea
where the momentum conservation is included as a delta function in the center of mass frame and $x\equiv\frac{m_{\rm DM}}{T}$ with $T$ being the DM temperature that is equal to the background temperature in kinetic equilibrium.  In this case, the thermal average is simplified to the integral for relative velocity, $|{\vec v}_1-{\vec v}_2|\equiv v$, as follows,
\bea
\langle \sigma v\rangle=\frac{x^{3/2}}{2\sqrt{\pi}} \int^\infty_0 dv \, v^2 (\sigma v)\, e^{-\frac{1}{4} x v^2}.
\eea

Suppose to take the velocity expansion of the $2\rightarrow 2$ cross section as
\be
(\sigma v)=\sum_{l=0 }^\infty\frac{a_l}{l!}\, (v^2)^l .
\ee
Due to the absence of a resonance, we get the thermal average simply as
\bea
\langle \sigma v\rangle&=&\frac{1}{2\sqrt{\pi}}\sum_{l=0 }^\infty 4^{l+1} \Gamma\Big(l+\frac{3}{2}\Big)\, \frac{a_l}{l!}\,x^{-l} \nonumber \\
&=& a_0+6a_1 x^{-1}+30a_2 x^{-2}+\cdots . \label{wimp1}
\eea
Thus, we have recovered the well known results for the thermal-averaged $2\rightarrow 2$ annihilation cross section \cite{gondolo}.

On the other hand, in the presence of a resonance $R$,  the $2\rightarrow 2$ annihilation cross section for ${\chi\chi\rightarrow R \rightarrow f{\bar f}}$ takes the following Breit-Wigner form,
\bea
(\sigma v)_R&=& \frac{32\pi}{m^2_R \beta_\chi} \frac{\gamma^2_R}{(\epsilon_R-\eta)^2+\gamma^2_R}\, {\rm Br}(R\rightarrow \chi\chi)\, {\rm Br}(R\rightarrow f{\bar f}) \nonumber \\
&\equiv &\sum_{l=0 }^\infty\frac{b_l}{l!}\, \eta^l \frac{\gamma^2_R}{(\epsilon_R-\eta)^2+\gamma^2_R}, \label{res22}
\eea
where $\beta_\chi$ is the DM velocity, and $\eta\equiv \frac{1}{4} v^2$,  $\epsilon_R\equiv\frac{m^2_R-4m^2_{\rm DM}}{4m^2_{\rm DM}}$ and $\gamma_R\equiv \frac{m_R \Gamma_R}{4m^2_{\rm DM}}$, with $m_R, \Gamma_R$ being the mass and width of the resonance. 
Then, we obtain the general result for the thermal average with a resonance as follows,
\bea
\langle\sigma v\rangle_R =2x^{3/2}\sqrt{\pi}\gamma_R\sum_{l=0 }^\infty\frac{b_l}{l!}\,F_l(z_R;x) \label{wimp2}
\eea
where $z_R\equiv \epsilon_R+i\gamma_R$ and
\bea
F_l(z_R;x) &=& {\rm Re}\bigg[\frac{i}{\pi}\int^\infty_0 \frac{\eta^{l+1/2} e^{-x \eta}\, d\eta}{z_R-\eta} \bigg]  \nonumber \\
&=& (-1)^l \frac{\partial^l}{\partial x^l}\, F_0(z_R,x).
\eea
Here, the generating integral is given by
\bea
F_0 (z_R;x)&=& {\rm Re}\bigg[\frac{i}{\pi}\int^\infty_0 \frac{\eta^{1/2} e^{-x \eta}\, d\eta}{z_R-\eta} \bigg] \nonumber \\
&=&{\rm Re}\Big[z^{1/2}_R e^{-x z_R} {\rm Erfc}(-ix^{1/2} z^{1/2}_R) \Big],
\eea
with the complementary error function being given by 
\be
{\rm Erfc}(a) \equiv \frac{2}{\sqrt{\pi}} \int^\infty_a e^{-t^2}\, dt. 
\ee
In particular, in the narrow width approximation with $\gamma_R\ll 1$, we get $F_0(z_R;x)\approx \epsilon^{1/2}_R e^{-x \epsilon_R}\theta(\epsilon_R)$ with $\theta(\epsilon_R)=1$ for $\epsilon_R>0$; $\theta(\epsilon_R)=0$ for $\epsilon_R<0$ and the thermal averaged cross section becomes
\bea
\langle\sigma v\rangle_R\approx 2\sqrt{\pi} \gamma_R\, \epsilon^{1/2}_R x^{3/2} e^{-x\epsilon_R} \theta(\epsilon_R) \sum_{l=0 }^\infty\frac{b_l}{l!}\,\epsilon^l_R.  \label{wimp3}
\eea
Thus, the averaged annihilation cross section becomes a step function in the narrow width approximation, being sensitive to the resonance mass \cite{gondolo,griest}. 

\begin{figure}
  \begin{center}
   \includegraphics[height=0.42\textwidth]{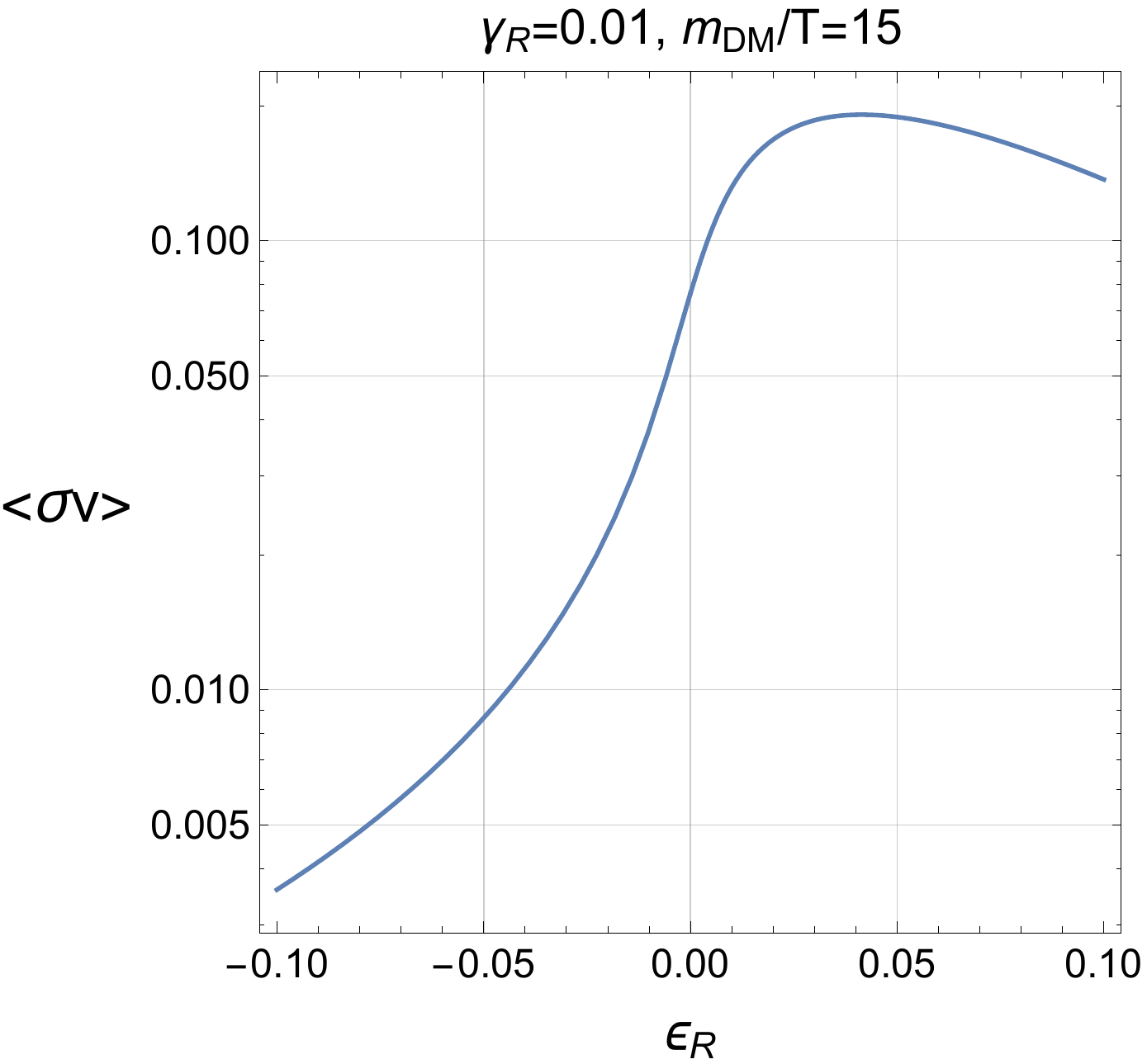}
      \includegraphics[height=0.42\textwidth]{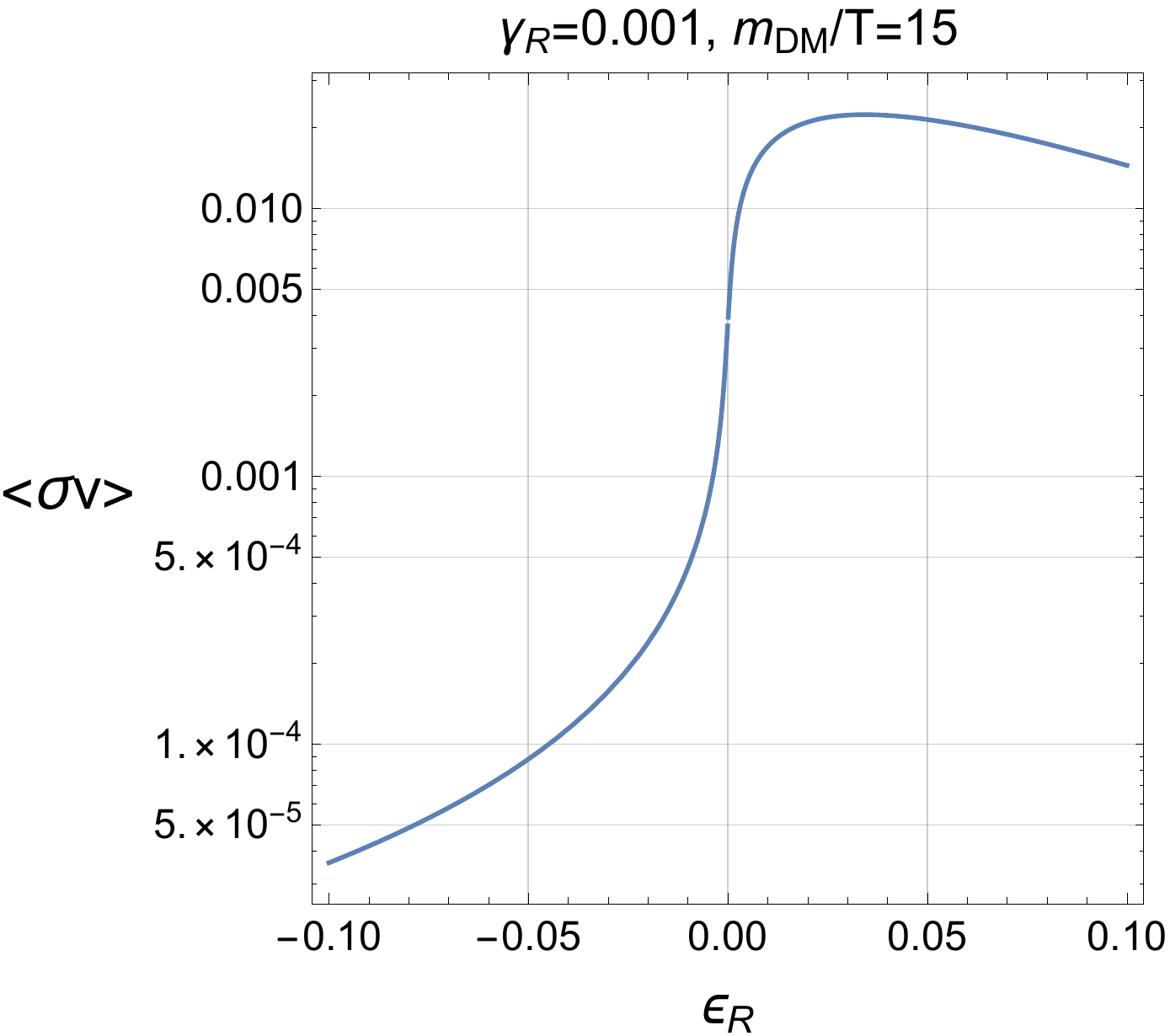}
   \end{center}
  \caption{Thermal-averaged $2\rightarrow 2$ annihilation cross section near resonance as a function of $\epsilon_R$. Here, the cross section is given in arbitrary unit, so only the relative ratio at different values of $\epsilon_R$ is important.}
  \label{wimp}
\end{figure}

In Fig.~\ref{wimp}, we show the exact results for the averaged annihilation cross section with $s$-wave overall factor in arbitrary unit as a function of $\epsilon_R$ for a fixed $\gamma_R$ and temperature, $T=\frac{m_{\rm DM}}{15}$.  In the limit of a narrow width, the averaged annihilation cross section is shown to be step-wise as in our approximate formula in eq.~(\ref{wimp3}).

\section{Thermal average for $3\rightarrow 2$ DM annihilations}

Assuming that three DM particles in the initial states have the same masses, $m_1=m_2=m_3\equiv m_{\rm DM}$, the thermal averaged $3\rightarrow 2$ cross section is given by
\bea
\langle \sigma v^2\rangle= \frac{\int d^3 v_1 d^3 v_2 d^3 v_3\, \delta^3({\vec v}_1+{\vec v}_2+{\vec v}_3)(\sigma v^2)\, e^{-\frac{1}{2}x(v^2_1+v^2_2+v^2_3)}}{\int d^3 v_1 d^3 v_2 d^3 v_3\, \delta^3({\vec v}_1+{\vec v}_2+{\vec v}_3) \,e^{-\frac{1}{2}x(v^2_1+v^2_2+v^2_3)}}.
\eea
We assumed that the spins of dark matter are averaged and summed  over initial and final states in $3\rightarrow 2$ processes. Then, the resulting velocity expansion of the $3\rightarrow 2$ cross section depends on the spin and parity of dark matter. For instance, in the case of fermionic SIMP, the initial states in the $3\rightarrow 2$ process can be all fermions as discussed in Ref.~\cite{coann} while the case of vector SIMP was discussed \cite{vsimp} or will be published elsewhere \cite{vsimp2}.

In the non-relativistic limit of dark matter, 
taking into account the Galilean symmetry and permutation symmetry between three initial DM particles, we can take the velocity expansion of the $3\rightarrow 2$ cross section as follows,
\bea
(\sigma v^2)= a_0 + a_1 (v^2_1+v^2_2+v^2_3) + a^{(1)}_2 (v^2_1+v^2_2+v^2_3)^2 + a^{(2)}_2 (v^4_1+v^4_2+v^4_3)+\cdots .
\eea
There appear more combinations of squared velocities at higher orders. 
We note that at the fourth order in velocities, an alternative basis can be choosen with $v^2_1 v^2_2 +v^2_2 v^2_3+v^2_3 v^2_1$ or $({\vec v}_1\cdot {\vec v}_2)^2+({\vec v}_2\cdot {\vec v}_3)^2+(\vec v_3\cdot {\vec v}_1)^2$, instead of $v^4_1+v^4_2+v^4_3$, whenever it is more convenient for thermal average \footnote{We note the following identities, $v^2_1 v^2_2+v^2_2 v^2_3+ v^2_3 v^2_1=\frac{1}{2}(v^2_1+v^2_2+v^2_3)^2-\frac{1}{2}(v^4_1+v^4_2+v^4_3)$, and $({\vec v}_1\cdot {\vec v}_2)^2+({\vec v}_2\cdot {\vec v}_3)^2+(\vec v_3\cdot {\vec v}_1)^2=v^4_1+v^4_2+v^4_3-\frac{1}{4}(v^2_1+v^2_2 +v^2_3)^2$ due to ${\vec v}_1+{\vec v}_2+{\vec v}_3=0$.}.

\subsection{Non-resonance}

The thermal average of velocity terms, given by a function of $v_1^2+v^2_2+v^2_3$, namely in an $SO(9)$ symmetric form, can be easily computed in a closed form as below. Thus, we first treat them separately and next consider general terms of the form, $(v^2_1)^n(v^2_2)^m(v^2_3)^l$.

First, we take the velocity expansion of the $3\rightarrow 2$ cross section in the following form with $SO(9)$ invariance,
\be
(\sigma v^2)= \sum_{l=0}^\infty \frac{a_l}{l!}\, \eta^l
\ee
with $\eta\equiv \frac{1}{2}(v^2_1+v^2_2+v^2_3)$.
Then, the corresponding thermal average is given by
\bea
\langle \sigma v^2\rangle&=& \frac{1}{2} x^3 \sum_{l=0}^\infty \frac{a_l}{l!}\,\int^\infty_0 d\eta\, \eta^{l+2} e^{-x\eta} \nonumber \\
&=& \frac{1}{2}  \sum_{l=0}^\infty  (l+1)(l+2) a_l \,x^{-l} \nonumber \\
&=&a_0+3 a_1 x ^{-1}+6 a_2 x^{-2}+\cdots .  \label{etan}
\eea
In most cases, the most important terms appear up to $p$-wave terms that are $SO(9)$ invariant, so the above result gives rise to a good approximation for the full average.  
But, if the $3\rightarrow 2$ cross section is velocity-suppressed, we need to take into account the precise form of higher order terms in the velocity expansion. 

There are cases where the leading terms in the velocity expansion are higher than $p$-wave, such as in the case with SIMP mesons which have leading $d$-wave terms.  
Thus, for more general velocity terms,  we need to do the velocity integrations  as
\bea
\langle(v^2_1)^n(v^2_2)^m(v^2_3)^l\rangle &=& \frac{3\sqrt{3} x^3}{\pi}\, \int^\infty_0 dv_1 v^2_1 \int^\infty_0 dv_2 v_2^2 (v^2_1)^n (v^2_2)^m \times \nonumber \\
&&\quad \times \int^{+1}_{-1} d\cos\theta_{12} (v^2_1+v^2_2+2v_1 v_2 \cos\theta_{12})^l \, e^{-x(v^2_1+v^2_2+v_1 v_2 \cos\theta_{12})} \nonumber \\
&\equiv & c_{nml} \, x^{-n-m-l}\,
\eea 
where $c_{nml}$ are constant coefficients depending on $(n,m,l)$. 
In the case with $l=0$, the above integration can be simplified to
\bea
\langle(v^2_1)^n(v^2_2)^m\rangle= \frac{3\sqrt{3} x^2}{\pi}\, \int^\infty_0 dv_1\, v^{2n+1} e^{-\frac{3}{4}x v^2_1}\int^\infty_{-\infty} dv'_2 \Big(v'_2+\frac{1}{2}v_1\Big)^{2m+1} e^{-x v^{\prime 2}_2}. 
\eea
These integrals can be calculated numerically and some of them with low $n, m, l$ are shown in Table 1. Other combinations with a fixed value of $n+m+l$ are not shown because they are the same as the one shown in Table 1 due to permutation symmetry between dark matter particles.

\begin{table}[ht]
\centering
\begin{tabular}{|c||c|c|c|c|c|c|}
\hline 
$(n,m,l)$ & $(1,0,0)$ & $(1,1,0)$ & $(2,0,0)$ & $(1,1,1)$ & $(2,1,0)$ & $(3,0,0)$   \\ [0.5ex]
\hline 
$c_{nml}$ & $2$ & $\frac{14}{3}$ & $\frac{20}{3}$ & $\frac{100}{9}$ & $\frac{160}{9}$ & $\frac{280}{9}$ 
 \\ [0.5ex]
\hline
\end{tabular}
\caption{Coefficients of thermal averaged velocity terms. }
\label{table:charges1}
\end{table}

Instead, taking $m=l=0$, we can perform the integral in a closed form as
\bea
\langle(v^2_1)^n\rangle=\Big(\frac{4}{3}\Big)^n \frac{\Gamma(n+\frac{3}{2})}{\Gamma(\frac{3}{2})}\, x^{-n}= \langle(v^2_2)^n\rangle= \langle(v^2_3)^n\rangle. \label{v1n}
\eea
In particular, using eqs.~(\ref{etan}) and (\ref{v1n}), we get the thermal average of $d$-wave terms as follows,
\bea
\langle\sigma v^2\rangle_{d{\rm-wave}}&=&a^{(1)}_2 \langle (v^2_1+v^2_2+v^2_3)^2\rangle + a^{(2)}_2 \langle(v^4_1+v^4_2+v^4_3)\rangle \nonumber \\
&=& (48 a^{(1)}_2+ 20a^{(2)}_2 ) x^{-2}. \label{dwave}
\eea
In most of examples for $3\rightarrow  2$ processes such as SIMP mesons, it would be sufficient to consider at most the $d$-wave terms for thermal average.

\subsection{Resonance}

In the presence of resonances near the center of mass energy of three initial DM particles, more care is needed in the process of thermal average. 
In the non-relativistic limit of dark matter, the $3\rightarrow 2$ cross section for $\chi\chi\chi\rightarrow R\rightarrow  \chi\chi$, before thermal average, takes a generalized Breit-Wigner form,
\bea
(\sigma v^2)_R&=&\frac{9\sqrt{5}}{2 \beta_\chi \Phi_3 \, m^3_R}\frac{\gamma^2_R}{(\epsilon_R-\frac{2}{3}\eta)^2+\gamma^2_R}\,{\rm Br}(R\rightarrow \chi\chi\chi) \, {\rm Br}(R\rightarrow \chi\chi)  \nonumber \\
&\equiv & b_R\,\frac{\gamma_R}{(\epsilon_R-\frac{2}{3}\eta)^2+\gamma^2_R} \label{res32}
\eea
where $\beta_\chi$ is the DM velocity in the two-body decay of the resonance, namely, $\beta_\chi\equiv \sqrt{1-4m^2_\chi/m^2_R}$, $\Phi_3$ is the phase space integral for the three-body decay of the resonance, $R\rightarrow\chi\chi\chi$, and  $\epsilon_R, \gamma_R$ are the counterparts for the $3\rightarrow 2$ resonance, given by $\epsilon_R\equiv\frac{m^2_R-9m^2_{\rm DM}}{9m^2_{\rm DM}}$ and $\gamma_R\equiv \frac{m_R \Gamma_R}{9m^2_{\rm DM}}$, with $m_R, \Gamma_R$ being the mass and width of the resonance.
We note that the three-body phase space integral $\Phi_3$ is proportional to $\epsilon^2_R m^2_R$ near resonance, so the three-body decay rate of the resonance is suppressed as compared to the two-body decay rate.

\begin{figure}
  \begin{center}
   \includegraphics[height=0.42\textwidth]{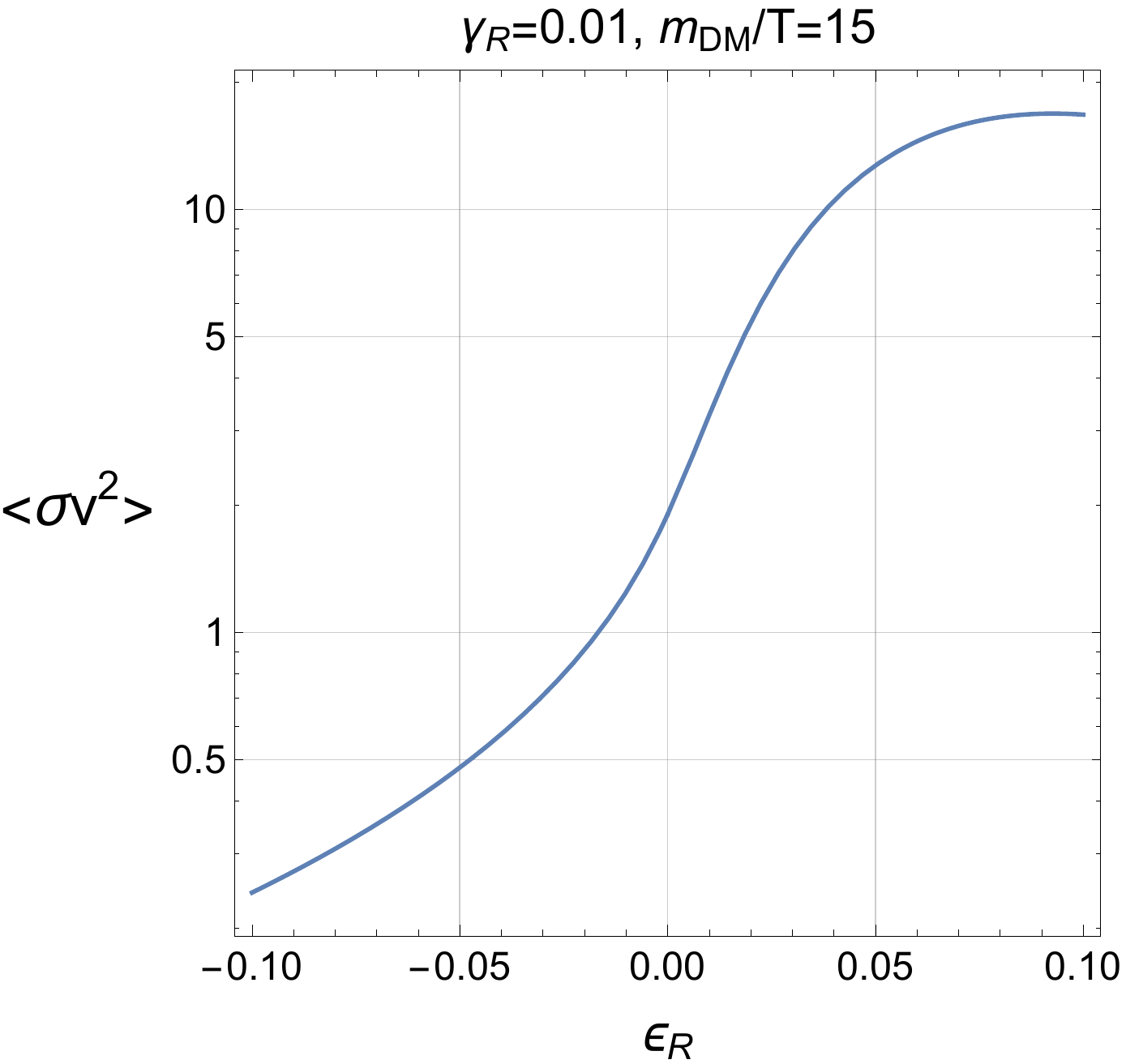}
      \includegraphics[height=0.42\textwidth]{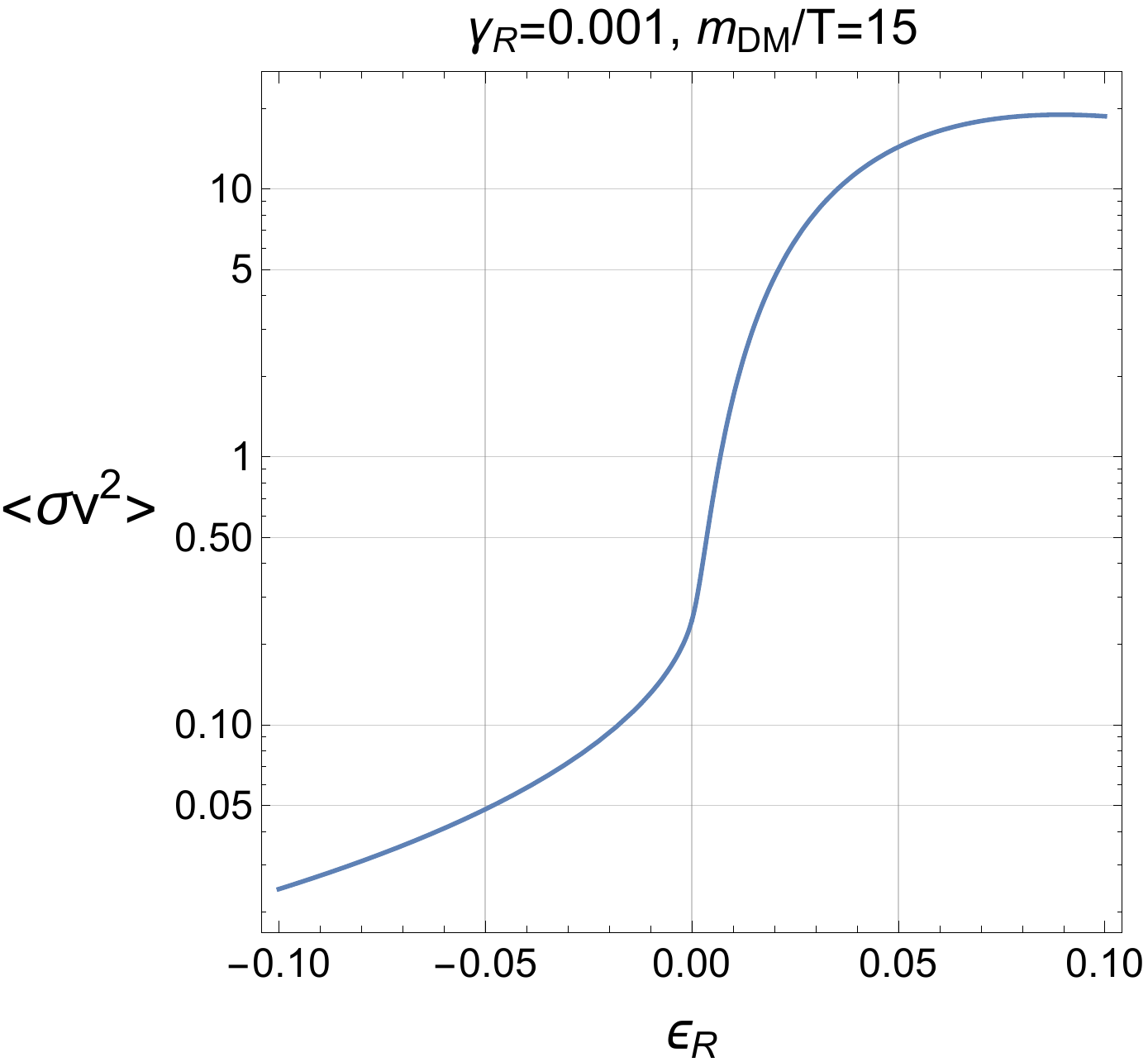}
   \end{center}
  \caption{Thermal-averaged $3\rightarrow 2$ annihilation cross section near resonance as a function of $\epsilon_R$. Here, the cross section is given in arbitrary unit, so only the relative ratio at different values of $\epsilon_R$ is important.}
  \label{simp}
\end{figure}

First, when the overall factor of the $3\rightarrow 2$ cross section is taken as a function of $\eta$ as $b_R=\sum_{l=0}^\infty \frac{b^{(l)}_R}{l!}\,\eta^l$,
the resulting thermal average is given by
\bea
\langle\sigma v^2 \rangle_R=\frac{3}{4}\pi x^3\sum_{l=0}^\infty \frac{b^{(l)}_R}{l!}\, G_l(z_R;x), \label{resonance}
\eea
where
\bea
G_l(z_R;x)&=& {\rm Re}\bigg[\frac{i}{\pi}\int^\infty_0 d\eta\, \frac{ \eta^{l+2} e^{-x\eta}}{\frac{3}{2}z_R-\eta} \bigg] \nonumber \\
&=&(-1)^l \frac{\partial^l}{\partial x^l}\, G_0(z_R;x). 
\eea
with $z_R\equiv \epsilon_R +i\gamma_R$. 
Here, the generating integral $G_0(z_R;x)$ can be written in a closed form as follows,
\bea
G_0(z_R;x)&=& {\rm Re}\bigg[\frac{i}{\pi}\int^\infty_0 d\eta\, \frac{ \eta^{2} e^{-x\eta}}{\frac{3}{2}z_R-\eta} \bigg]  \nonumber \\
&=&\frac{3}{2\pi}\,\frac{\gamma_R}{x}-\frac{9}{4\pi} {\rm Re}\Big[ i\, e^{-\frac{3}{2}x z_R}z^2_R\Big(\Gamma\Big(0,-\frac{3}{2}xz_R\Big)+\ln\Big(-\frac{1}{z_R}\Big)+\ln(-z_R) \Big) \Big]
\eea
where the incomplete gamma function being is given by
\be
\Gamma(0,a)\equiv \int^\infty_a\frac{e^{-t}}{t}\,dt.
\ee

For narrow width approximation with $\gamma_R\ll 1$, we get $G_0(z_R; x)\approx \frac{9}{4} \epsilon^2_R e^{-\frac{3}{2} x \epsilon_R}\theta(\epsilon_R)$  and the thermal averaged cross section becomes
\bea
\langle\sigma v^2 \rangle_R\approx \frac{27}{16}\pi\epsilon^2_R  x^3e^{-\frac{3}{2} x \epsilon_R}\theta(\epsilon_R)  \sum_{l=0}^\infty \frac{b^{(l)}_R}{l!}\, \Big(\frac{3}{2}\Big)^l \epsilon^l_R. \label{nwa}
\eea
We find that the averaged cross section in the SIMP case is more sensitive to the resonance mass through $\epsilon^2_R$  than in the WIMP case where the averaged cross section is proportional to $\epsilon^{1/2}_R$ in eq.~(\ref{wimp3}).  This is due to the fact that the phase space in the velocity average for three initial DM particles takes a higher power in DM velocity so it becomes more sensitive to the pole of the resonance.   

In Fig.~\ref{simp}, we depict the analytic results for thermal-averaged $3\rightarrow 2$ annihilation cross section with $s$-wave overall factor in arbitrary unit as a function of $\epsilon_R$ for a fixed $\gamma_R$ and temperature, $T=\frac{m_{\rm DM}}{15}$. Similarly to the WIMP case, the result is sensitive to the mass of the resonance and it becomes step-wise in the limit of a narrow width.

\section{Boltzmann equations for dark matter}

We use the general results on thermal averages in the previous section to solve the Boltzmann equations for the relic density of WIMP or SIMP dark matter.

\subsection{Boltzmann equation for WIMP}

The Boltzmann equation for WIMP dark matter is given by
\be
\frac{d n_{\rm DM}}{dt}+3H n_{\rm DM}= -\langle\sigma v\rangle (n^2_{\rm DM}- (n^{\rm eq}_{\rm DM})^2). 
\ee
Then, the above equation can be rewritten in terms of the relic abundance of dark matter, $Y_{\rm DM}=n_{\rm DM}/s$, as follows,
\bea
\frac{dY_{\rm DM}}{dx}= -\lambda x^{-2}  \langle\sigma v\rangle  \Big(Y^2_{\rm DM} - (Y^{\rm eq}_{\rm DM})^2 \Big)
\eea
where $\lambda\equiv s(m_{\rm DM})/H(m_{\rm DM})$  with $s(m_{\rm DM})=\frac{2\pi^2}{45} g_{*s} m^3_{\rm DM}$  and $1/H(m_{\rm DM})=3.02 g^{-1/2}_* \frac{M_P}{m^2_{\rm DM}}$. 
Therefore, we obtain the solution to the Boltzmann equation as
\bea
Y_{\rm DM}(\infty)\approx \bigg(\lambda J(x_f) \bigg)^{-1}.
\eea
with
\be
J(x_f)\equiv \int^\infty_{x_f} dx \, x^{-2} \langle\sigma v\rangle.
\ee
Here, $x_f=m_{\rm DM}/T_f$ with $T_f$ being the freeze-out temperature.
In the case without a resonance,  when $\langle\sigma v\rangle=a_l x^{-l}$ from eq.~(\ref{wimp1}), the $J$ factor becomes
\be
J(x_f)= \frac{a_l}{l+1}\,x^{-l-1}_f. 
\ee
As a result, the relic density of WIMP dark matter is given by
\bea
\Omega_{\rm WIMP} h^2&=&\frac{m_{\rm DM}Y_{\rm DM}(\infty) s_0}{3M_P^2 H^2_0/h^2} \nonumber \\
&=&\frac{8.53\times 10^{-11}\,{\rm GeV}^{-2}}{g^{1/2}_*J(x_f)}.
\eea

\begin{figure}
  \begin{center}
   \includegraphics[height=0.42\textwidth]{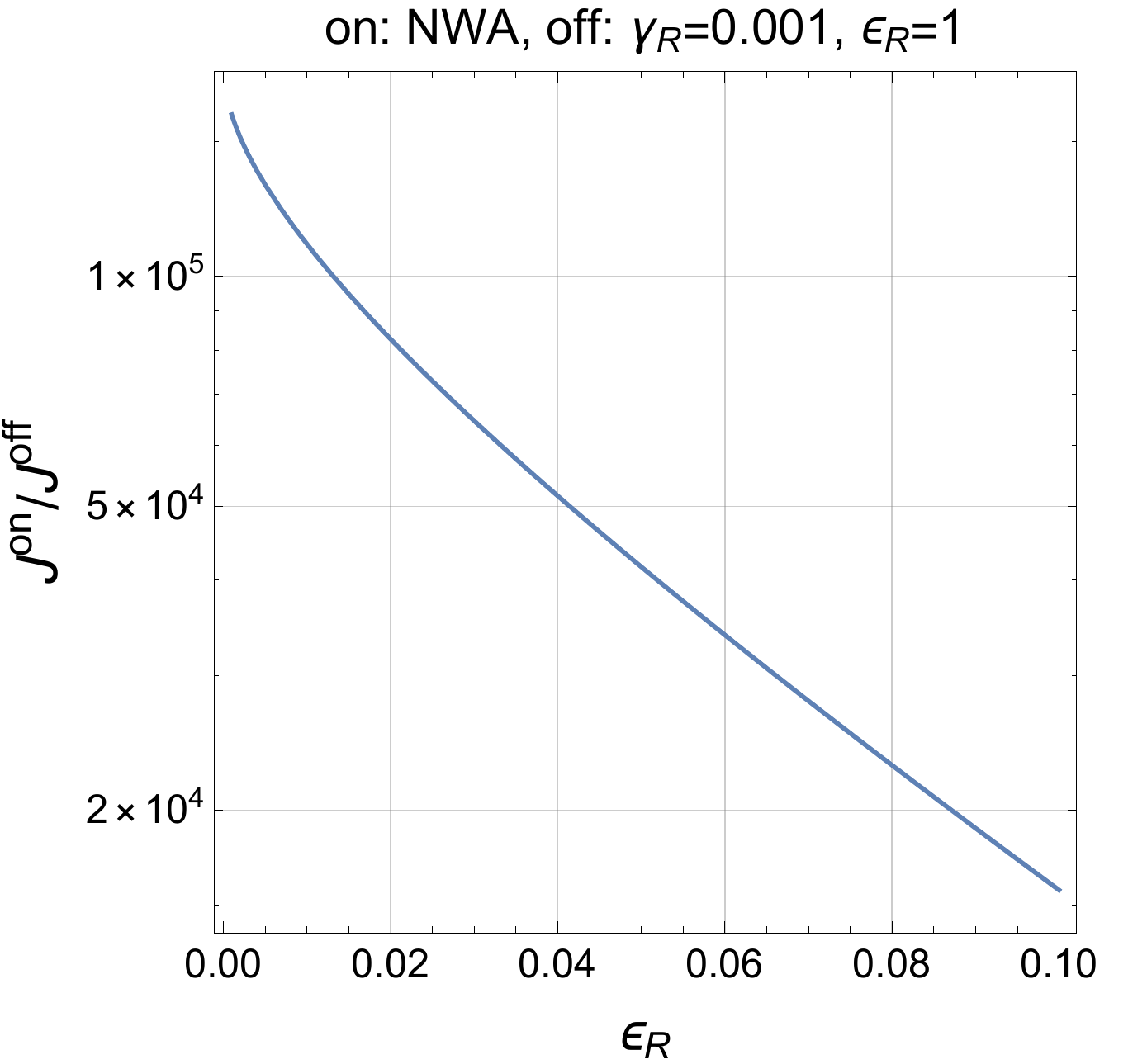}
      \includegraphics[height=0.42\textwidth]{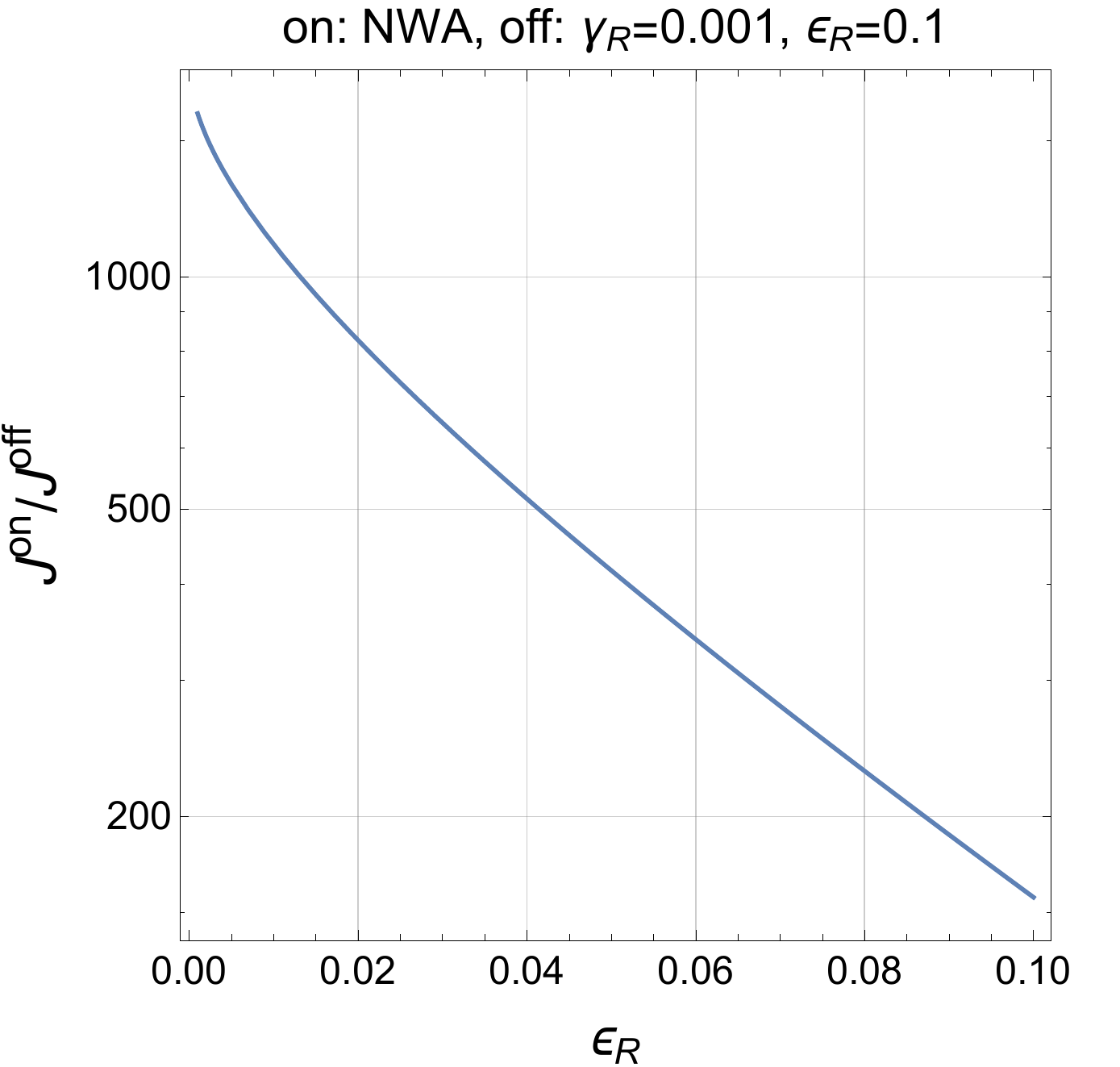}  \vspace{3mm} \\
        \includegraphics[height=0.42\textwidth]{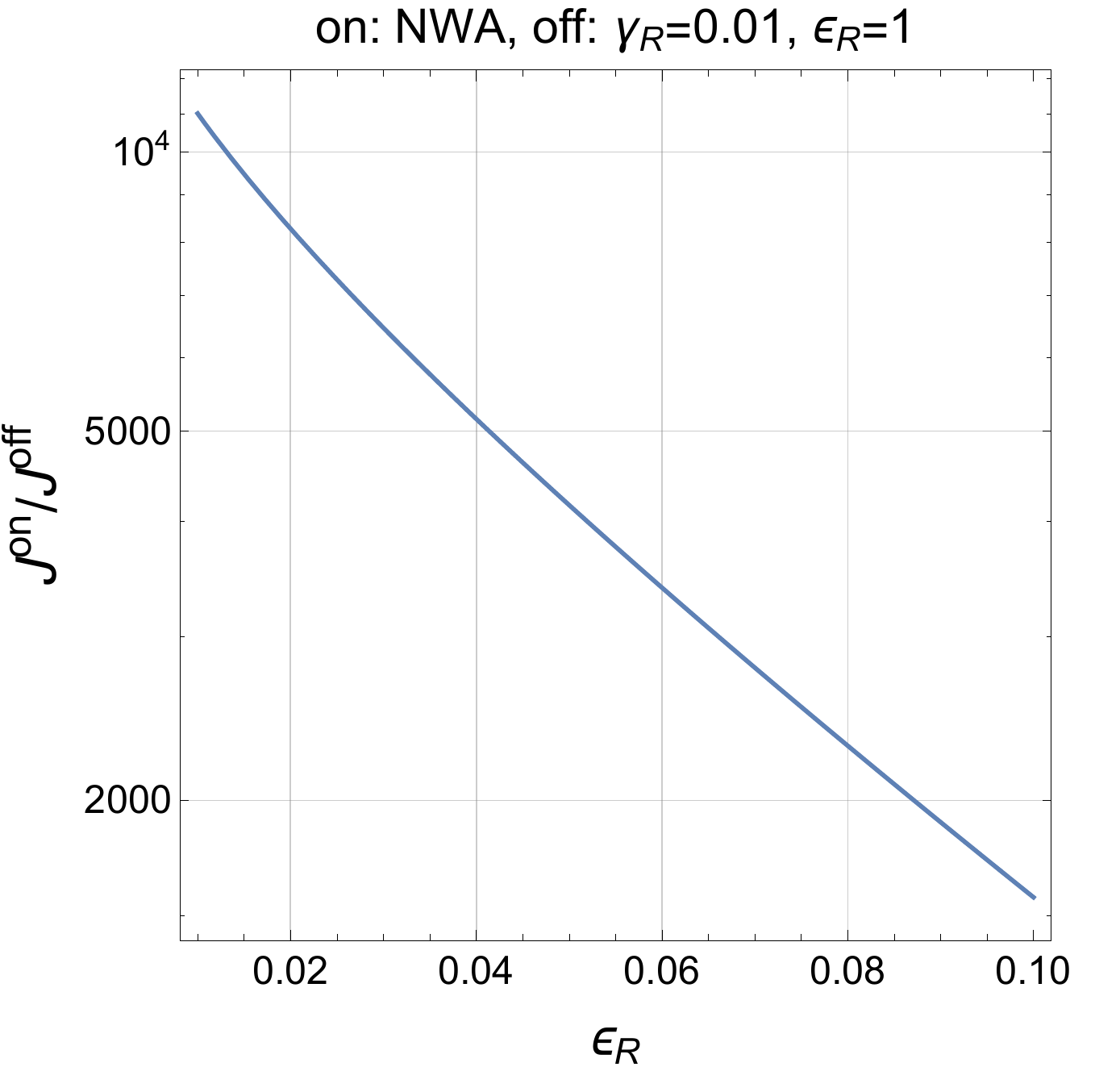}
      \includegraphics[height=0.42\textwidth]{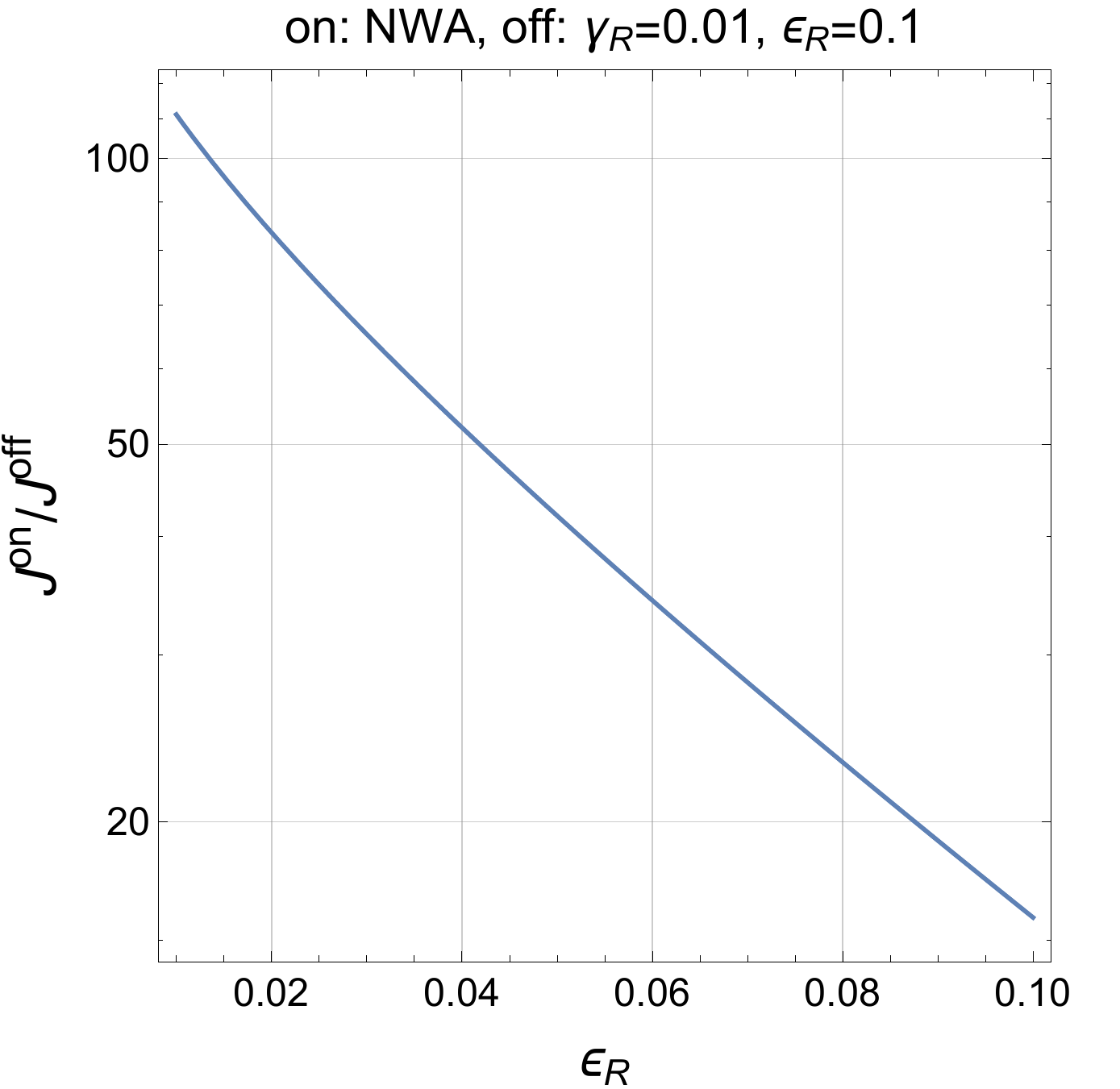}
   \end{center}
  \caption{$J$-factors on vs off-resonance for WIMP. Narrow Width Approximation(NWA) for the resonance is assumed and $\gamma_R, \epsilon_R$ are related to the resonance width and the amount of off-resonance as defined below eq.~(\ref{res22}).}
  \label{wimp-J}
\end{figure}

In the case with a resonance having a narrow width with $\epsilon_R>0$, when $\langle\sigma v\rangle=b_l \gamma_R\, \epsilon^{l+\frac{1}{2}}_R x^{3/2} e^{-x\epsilon_R}$ from eq.~(\ref{wimp3}), the $J$ factor becomes
\bea
J(x_f)&=&b_l \gamma_R\, \epsilon^{l+\frac{1}{2}}_R \int^\infty_{x_F} dx \, x^{-1/2} e^{-x\epsilon_R} \nonumber \\
&=&b_l \sqrt{\pi}\gamma_R\, \epsilon^{l}_R \,{\rm Erfc}(x^{1/2}_f\epsilon^{1/2}_R).
\eea
In Fig.~\ref{wimp-J}, we draw the ratio of $J$-factors for the $2\rightarrow 2$ annihilation cross section with $s$-wave overall factor at on- and off-resonance as a function of $\epsilon_R$. Thus, the large enhancement of the thermal-averaged cross section stands out in the $J$-factors, helping reducing the relic density to a right value without a large coupling. 
We note that the ratio of $J$-factors changes by order of magnitude, depending on $\epsilon_R$ below $0.1$.

\subsection{Boltzmann equation for SIMP}

The Boltzmann equation for SIMP dark matter is given by
\be
\frac{d n_{\rm DM}}{dt}+3H n_{\rm DM}= -\langle\sigma v^2\rangle (n^3_{\rm DM}- n^{\rm eq}_{\rm DM} n^2_{\rm DM}). 
\ee
Similarly as in the WIMP case,  we rewrite the above equation for the relic abundance of dark matter, $Y_{\rm DM}=n_{\rm DM}/s$, as follows,
\bea
\frac{dY_{\rm DM}}{dx}= -\rho x^{-5}  \langle\sigma v^2\rangle  \Big(Y^3_{\rm DM} - Y^{\rm eq}_{\rm DM} Y^2_{\rm DM}\Big)
\eea
where $\rho\equiv s^2(m_{\rm DM})/H(m_{\rm DM})$. 
Therefore, we obtain the solution to the Boltzmann equation as
\bea
Y_{\rm DM}(\infty)\approx \bigg(2\rho K(x_f) \bigg)^{-1/2}.
\eea
with
\be
K(x_f)\equiv \int^\infty_{x_f} dx \, x^{-5} \langle\sigma v^2\rangle.
\ee
As a result, the relic density of SIMP dark matter is given by
\bea
\Omega_{\rm SIMP} h^2&=&\frac{m_{\rm DM}Y_{\rm DM}(\infty) s_0}{3M_P^2 H^2_0/h^2} \nonumber \\
&=&\frac{1.05\times 10^{-10}\,{\rm GeV}^{-2}}{g^{3/4}_*m_{\rm DM} (K(x_f)/M_P)^{1/2} }.
\eea

\begin{figure}
  \begin{center}
   \includegraphics[height=0.42\textwidth]{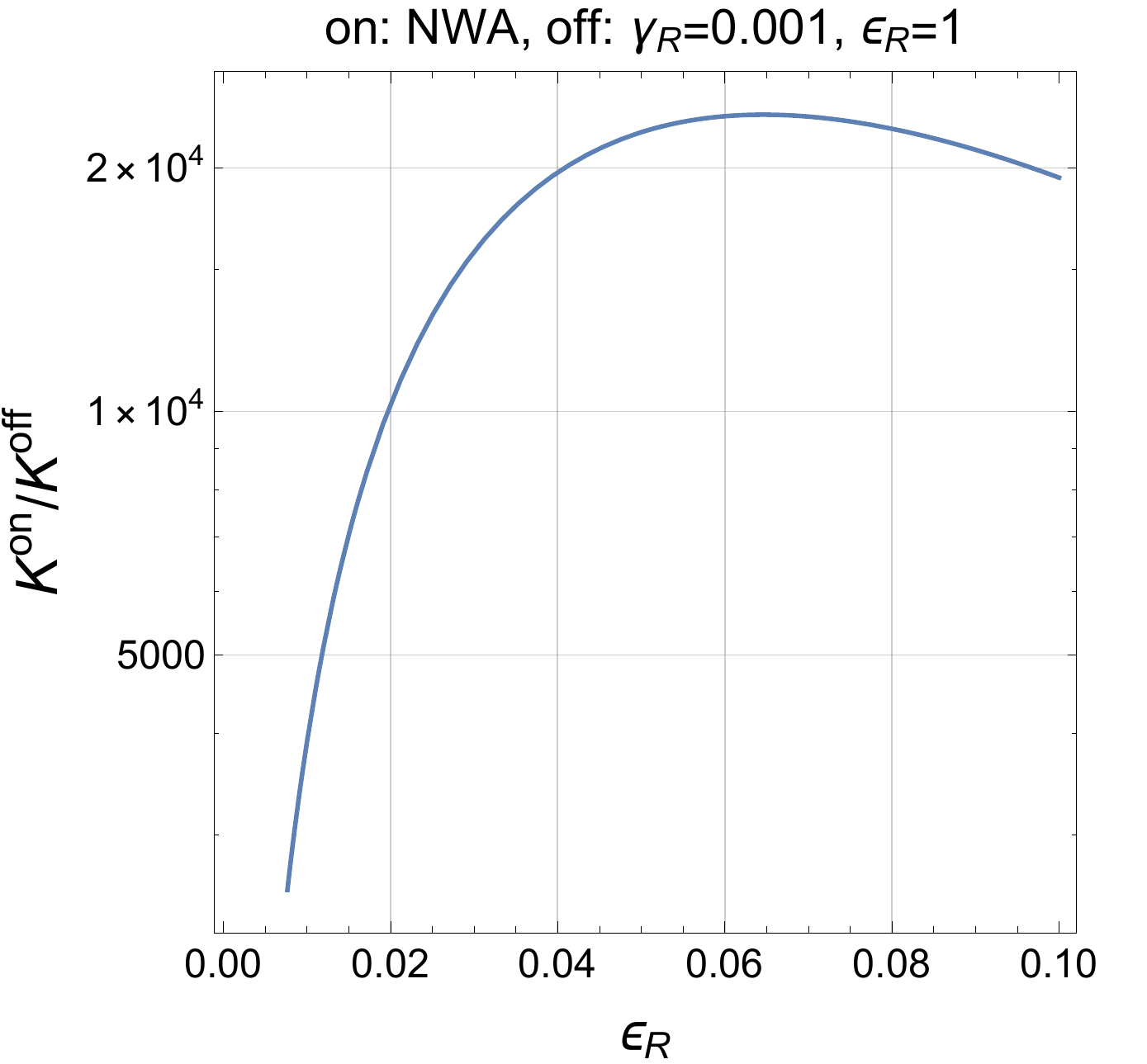}
      \includegraphics[height=0.42\textwidth]{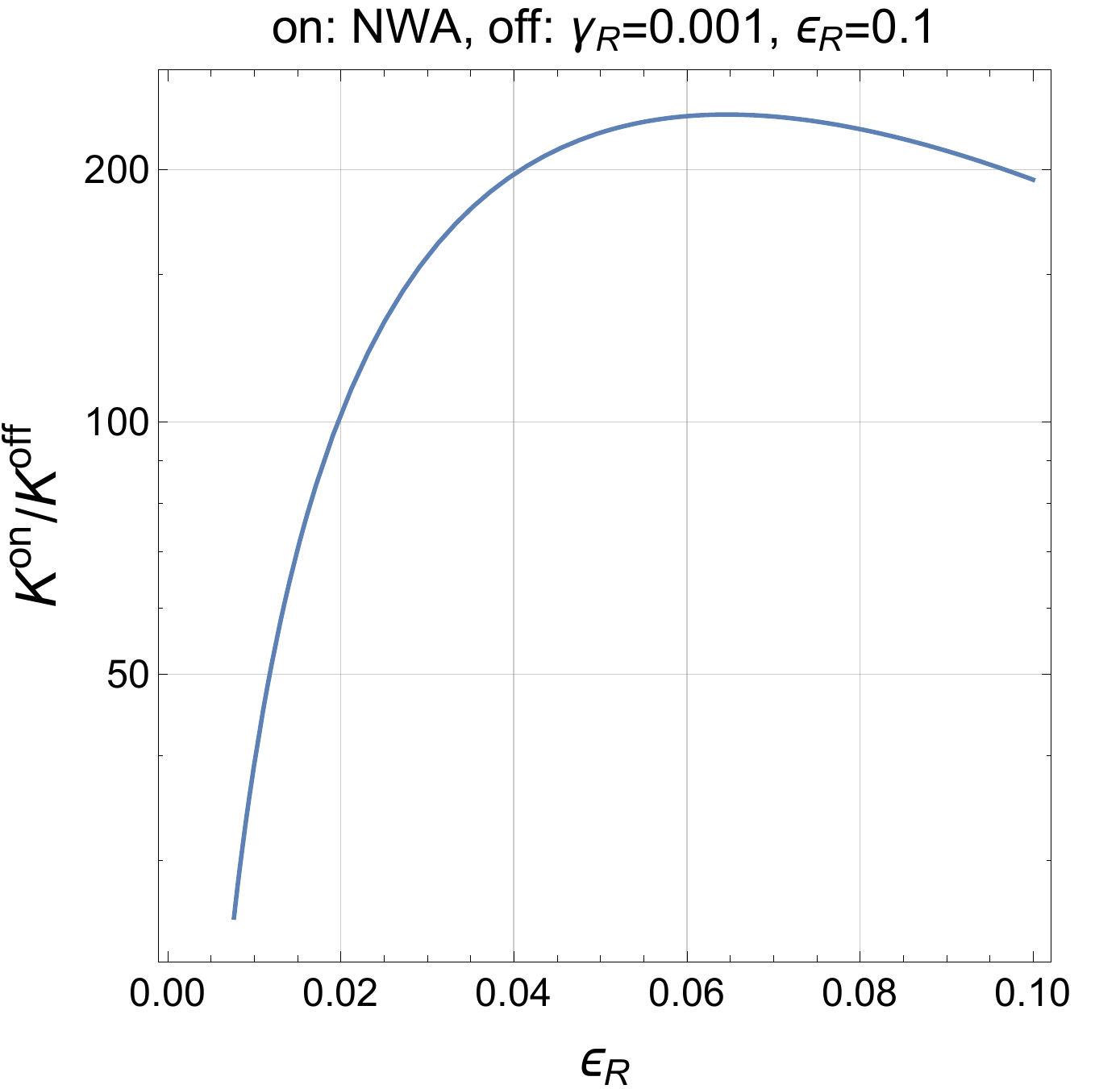}  \vspace{3mm} \\
        \includegraphics[height=0.42\textwidth]{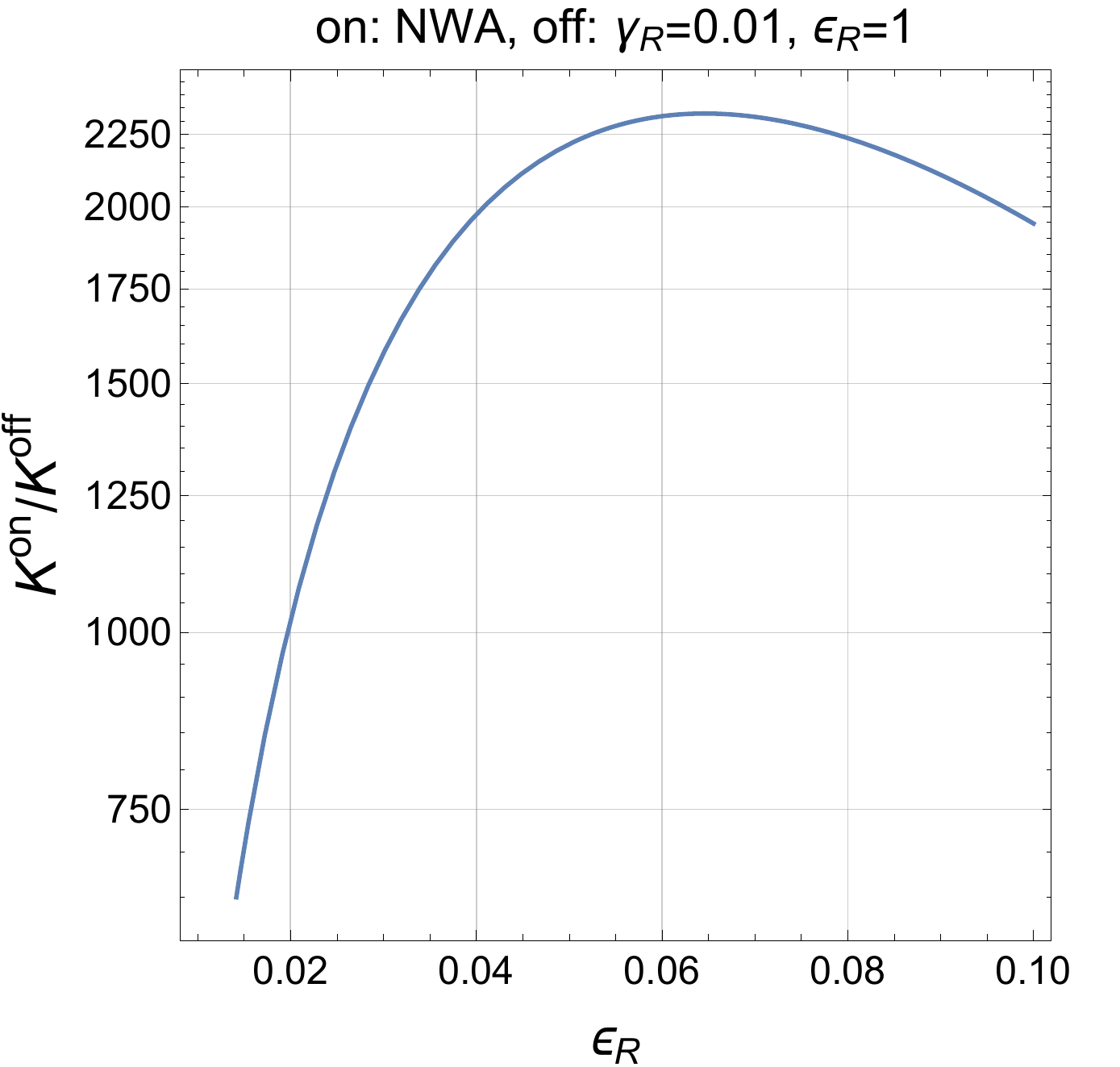}
      \includegraphics[height=0.42\textwidth]{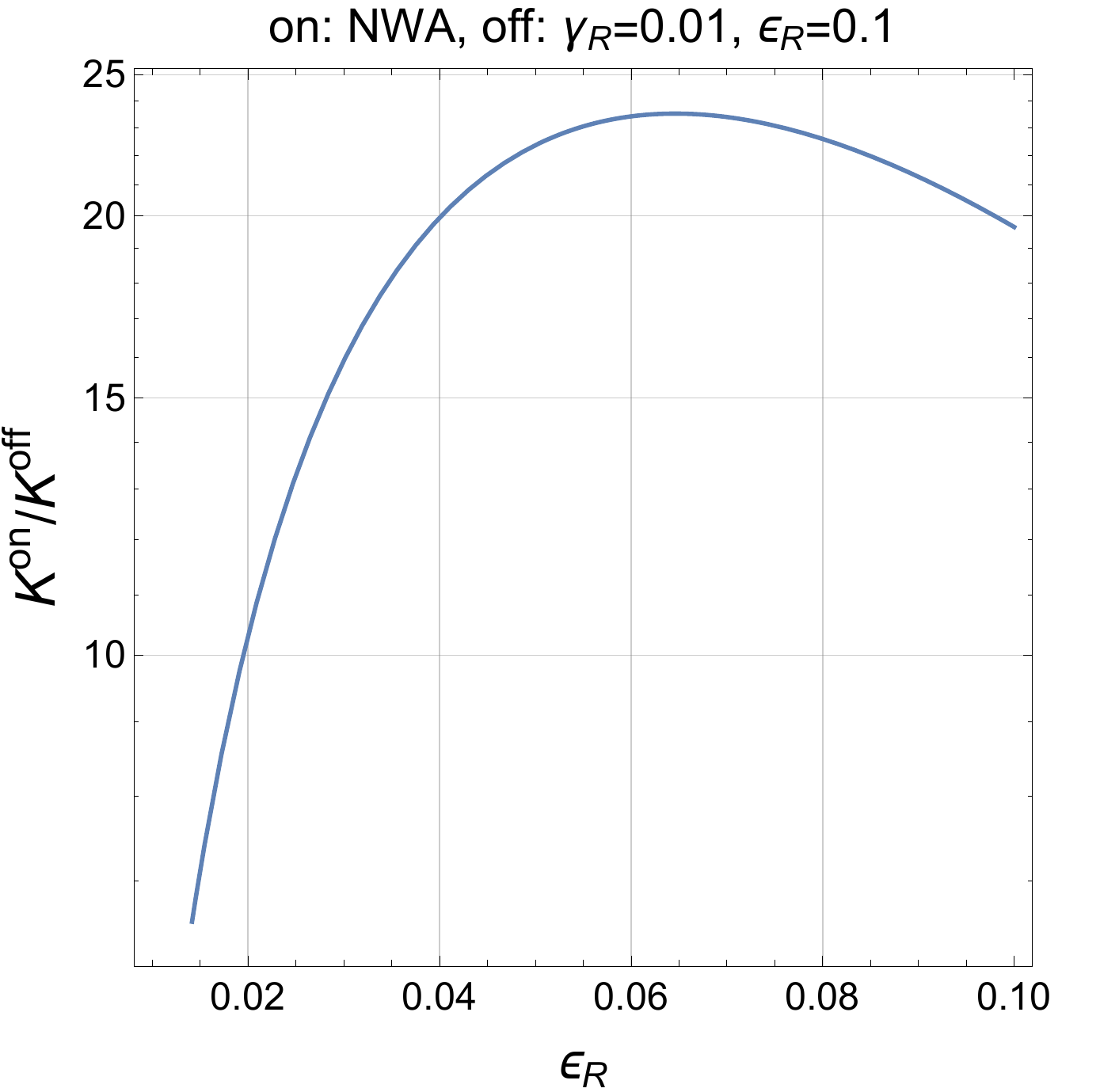}
   \end{center}
  \caption{$K$-factors on vs off-resonance for SIMP. Narrow Width Approximation(NWA) for the resonance is assumed and $\gamma_R, \epsilon_R$ are related to the resonance width and the amount of off-resonance as defined below eq.~(\ref{res32}).  } 
  \label{simp-K}
\end{figure}

In the case without a resonance,  when $\langle\sigma v^2\rangle=a_l x^{-l}$ from eq.~(\ref{etan}), the $K$ factor becomes
\be
K(x_f)= \frac{a_l}{l+4}\,x^{-l-4}_f. 
\ee
In the case with a resonance having a narrow width with $\epsilon_R>0$, when $\langle\sigma v^2\rangle=b_l\, \epsilon^{l+2}_R x^3 e^{-\frac{3}{2}x\epsilon_R}$ from eq.~(\ref{nwa}), the $K$ factor becomes
\bea
K(x_f)&=&b_l \, \epsilon^{l+2}_R \int^\infty_{x_F} dx \, x^{-2} e^{-\frac{3}{2}x\epsilon_R} \nonumber \\
&=&b_l\, \epsilon^{l+2}_R x^{-1}_f \bigg( e^{-\frac{3}{2}x_f\epsilon_R}-\frac{3}{2}x_f \epsilon_R \Gamma(0,\frac{3}{2} x_f\epsilon_R) \bigg).
\eea
As a result, we find that the $K$-factor has a different dependence on $\epsilon_R$ from the one of the $J$-factor in the previous section, due to the fact that the phase space in the velocity average for the SIMP case is more sensitive to $\epsilon_R$ than for the WIMP case. 
In Fig.~\ref{simp-K}, we depict the ratio of $K$-factors for the $3\rightarrow 2$ annihilation cross section with $s$-wave overall factor at off- and on-resonance as a function of $\epsilon_R$. 
Thus, we find that the $K$-factor becomes suppressed at small $\epsilon_R$ unlike the WIMP case while there is an optimal value of $\epsilon_R$ for which the $K$-factor is maximized.

\section{Benchmark models for SIMP dark matter}

In this section, we discuss some benchmark models for SIMP dark matter, with or without a resonance.
We first consider a complex scalar dark matter in models with discrete gauge symmetries and then dark mesons in models with hidden non-abelian gauge symmetries.

\subsection{SIMP dark matter with discrete gauge symmetries}

We consider discrete symmetries as remnants of a dark local $U(1)$ after it is spontaneously broken by a Higgs mechanism. Then, the $3\rightarrow 2$ processes appear with dark Higgs resonance $h'$ for the $Z_3$ case \cite{z3simp} and with extra scalar resonance $S$  for the $Z_5$ case \cite{z5simp}. Dark matter is a complex scalar $\chi$ with $q_\chi=+1$ in both cases or another complex scalar $S$ with $q_S=+3$ in the $Z_5$ case. 
In both cases, the $3\rightarrow 2$ processes are $s$-wave so our previous discussion in Section 2.2 for the thermal average of the $SO(9)$ invariant velocity expansion applies. 

After a dark local $U(1)$ is broken into a discrete symmetry $Z_n$ due to a VEV of a charged scalar $\phi$ with $q_\phi=n$, the relevant interaction terms for SIMP dark matter in the dark sector are given as follows \cite{z3simp,z5simp},
\bea
Z_3&:& {\cal L}_{Z_3}=- \kappa (v'+h') \chi^3+{\rm h.c.}-\lambda_\chi |\chi|^4-\frac{1}{2}\lambda_{\phi\chi} (v'+ h')^2 |\chi|^2,    \\
Z_5&:& {\cal L}_{Z_5}= -\lambda_1 v' S^2 \chi^\dagger -\lambda_2 v' S \chi^2 -\lambda_3 S^\dagger \chi^3 +{\rm h.c}.
\eea
Here, $v'$ is the VEV of a dark Higgs, which is expanded as $\phi=(v'+h')/\sqrt{2}$. 
Moreover, the dark photon $Z'$ gets mass of $m_{Z'}=3g_D v'$ or $5g_D v'$ in the $Z_3$ or $Z_5$ cases.  The resonance poles for $3\rightarrow 2$ processes appear at $m_{h'}=3m_\chi$ in the $Z_3$ case and $m_S=3m_\chi$ or $m_\chi=3m_S$ in the $Z_5$ case. 
For the $3\rightarrow 2$ dominance, we need to suppress the $2\rightarrow 2$ annihilations in the dark sector, requiring that $m_{Z'}, m_{h'}>m_\chi$.

\begin{figure}
  \begin{center}
   \includegraphics[height=0.42\textwidth]{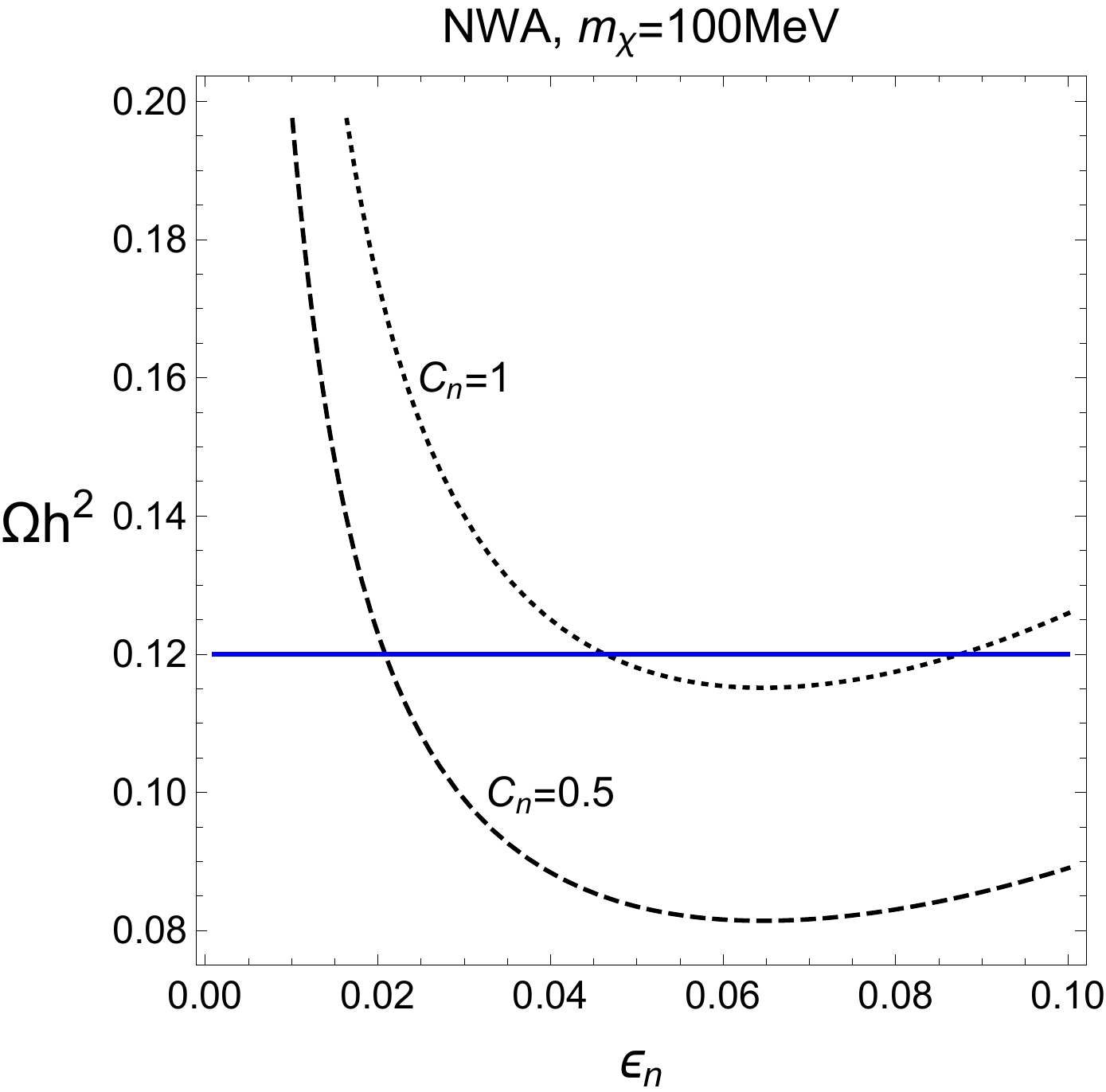}
      \includegraphics[height=0.42\textwidth]{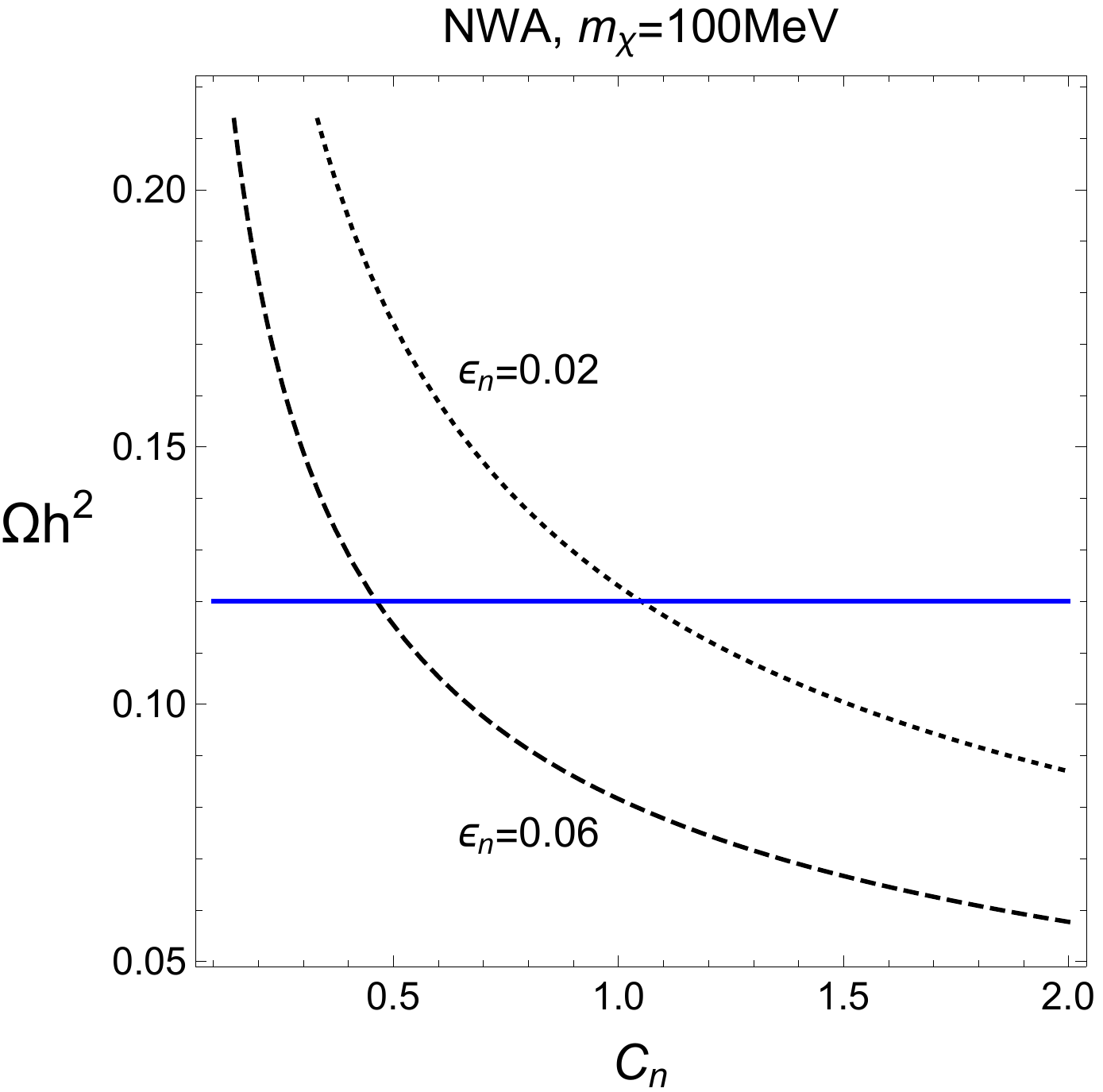}  \vspace{3mm} \\
        \includegraphics[height=0.42\textwidth]{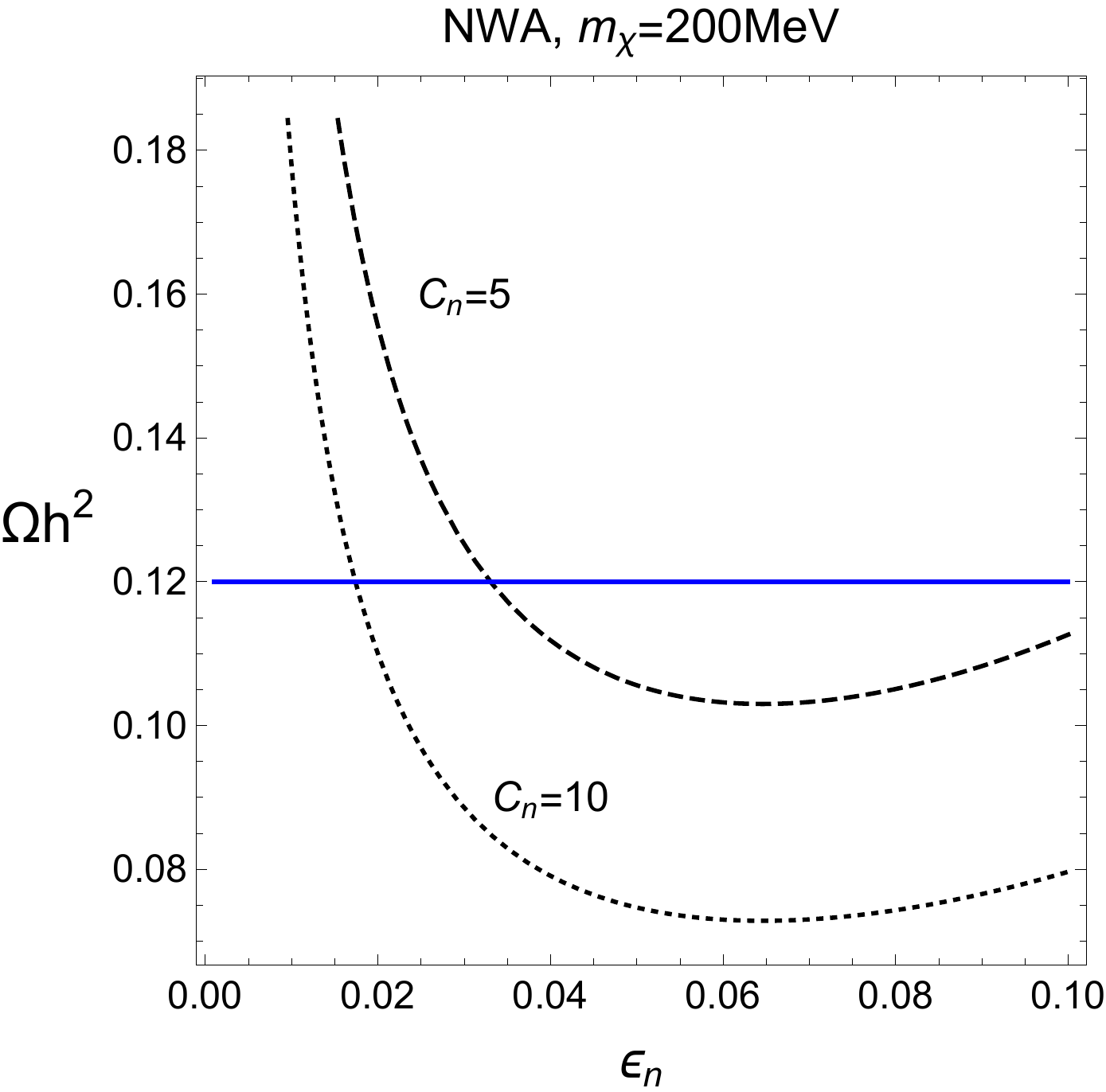}
      \includegraphics[height=0.42\textwidth]{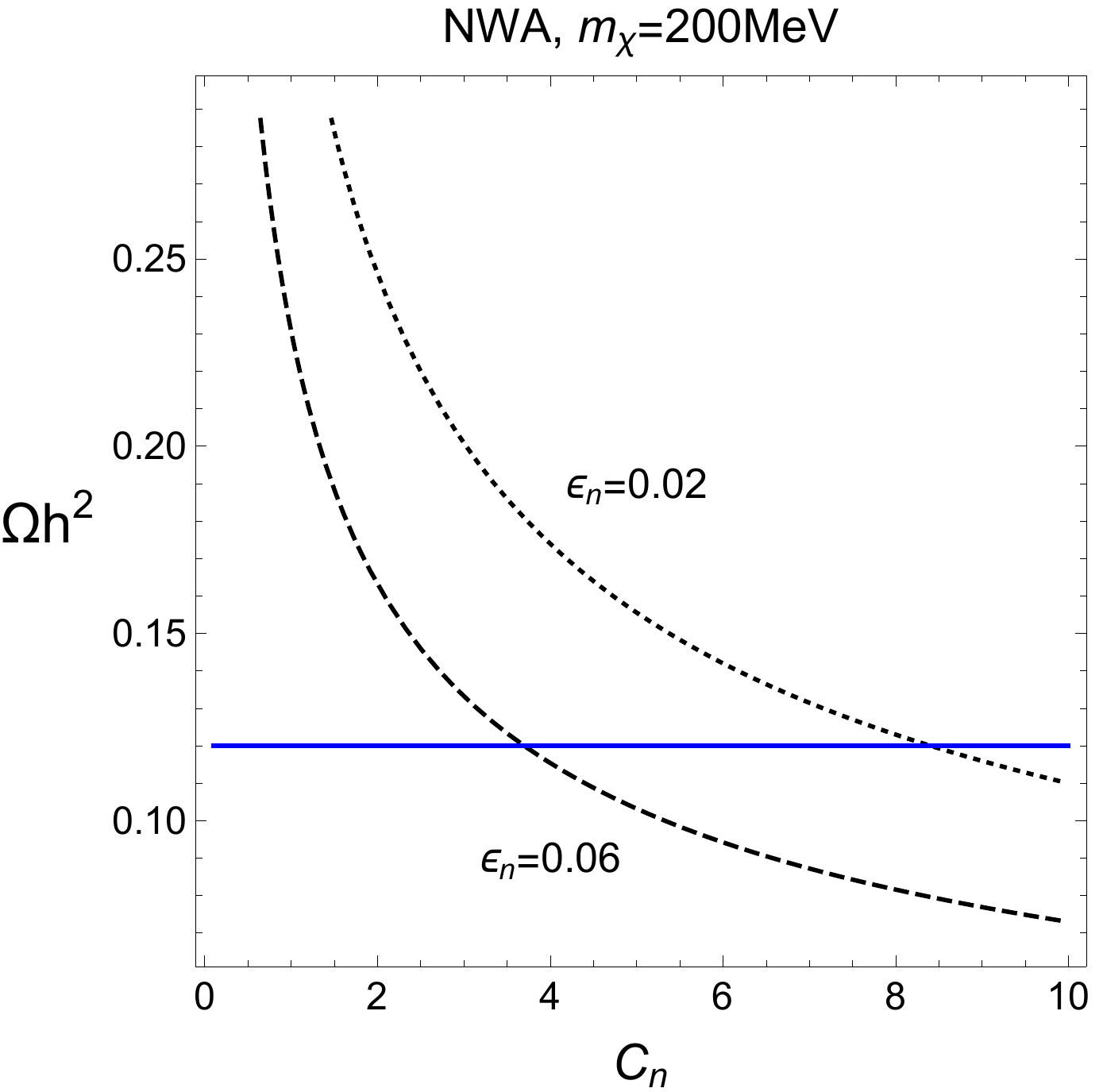}
   \end{center}
  \caption{Relic density as a function of $\epsilon_n$ ($C_n$) in the $s$-wave models in the left (right) panels. Blue solid line corresponds to the central value of the relic density by Planck. The results are shown in the narrow width approximation. }
  \label{swave}
\end{figure}

In the non-relativistic limit of dark matter, the $3\rightarrow 2$ annihilation cross sections with a resonance for discrete gauge symmetries take the form, 
\bea
(\sigma v^2)_{Z_n}= C_n \, \frac{\gamma_n}{(\epsilon_n-\frac{2}{3}\eta)^2+\gamma^2_n}
\eea
where $\epsilon_n=\frac{m^2_{n}-9m^2_\chi}{9m^2_\chi}$ and $\gamma_{n}=\frac{m_{n} \Gamma_n}{9m^2_\chi}$ with $m_3=m_{h'}$ and $m_5=m_{S}$ or $m_\chi$, and $C_n$ is given by
\bea
C_3&=& \frac{\sqrt{5} \kappa^2}{12{\beta}_\chi m^5_\chi} \Big(1+\frac{\lambda_{\phi\chi}v^{\prime 2}}{m^2_\chi}\Big)^2, \\
C^\chi_5&=& \frac{\sqrt{5}}{12 \beta'_\chi m^5_\chi} \Big(\lambda_3+\frac{2\lambda_1\lambda_2 v^{\prime 2}}{4m^2_\chi-m^2_S} \Big)^2, \\
C^S_5 &=& \frac{\sqrt{5}}{3 {\beta}_S m^5_S} \frac{\lambda^2_1 \lambda^2_2 v^{\prime 4}}{ (4m^2_S-m^2_\chi)^2}.
\eea
Here, $C^{\chi,S}_5$ denote the coefficients for $\chi$ and $S$ SIMP dark matters in $Z_5$ models, respectively, and  ${\beta}_\chi\equiv\sqrt{1-4m^2_\chi/m^2_{h'}}$, ${\beta}'_\chi\equiv\sqrt{1-4m^2_\chi/m^2_S}$, and $\beta_S\equiv \sqrt{1-4m^2_{S}/m^2_\chi}$.
 The width of the resonance is approximated by the partial decay width of the two-body decay mode, $h'\rightarrow \chi \chi^*$ in the $Z_3$ case and $S\rightarrow \chi^* \chi^*$ or $\chi\rightarrow SS$ in the $Z_5$ case, as follows,
\bea
\Gamma_{h'}&=& \frac{\lambda^2_{\phi\chi} v^{\prime 2}}{16\pi m_{h'}}\sqrt{1-\frac{4m^2_\chi}{m^2_{h'}}}, \\
\Gamma_S &=& \frac{\lambda^2_2 v^{\prime 2}}{8\pi m_S}\sqrt{1-\frac{4m^2_\chi}{m^2_S}}, \\
\Gamma_\chi&=& \frac{\lambda^2_1 v^{\prime 2}}{8\pi m_\chi} \sqrt{1-\frac{4m^2_S}{m^2_\chi}}.
\eea
We note that there are also three-body decay modes of the resonance in both $Z_3$ and $Z_5$ models, but the corresponding decay rates are suppressed by extra phase space, roughly by $\epsilon^2_n/(4\pi^2)$ for a constant squared decay amplitude, as compared to the two-body decay rates. Therefore, near the resonance with $\epsilon_n\lesssim 0.1$, the three-body decay contributions to the total decay rate of the resonance can be ignored.

\begin{figure}
  \begin{center}
   \includegraphics[height=0.52\textwidth]{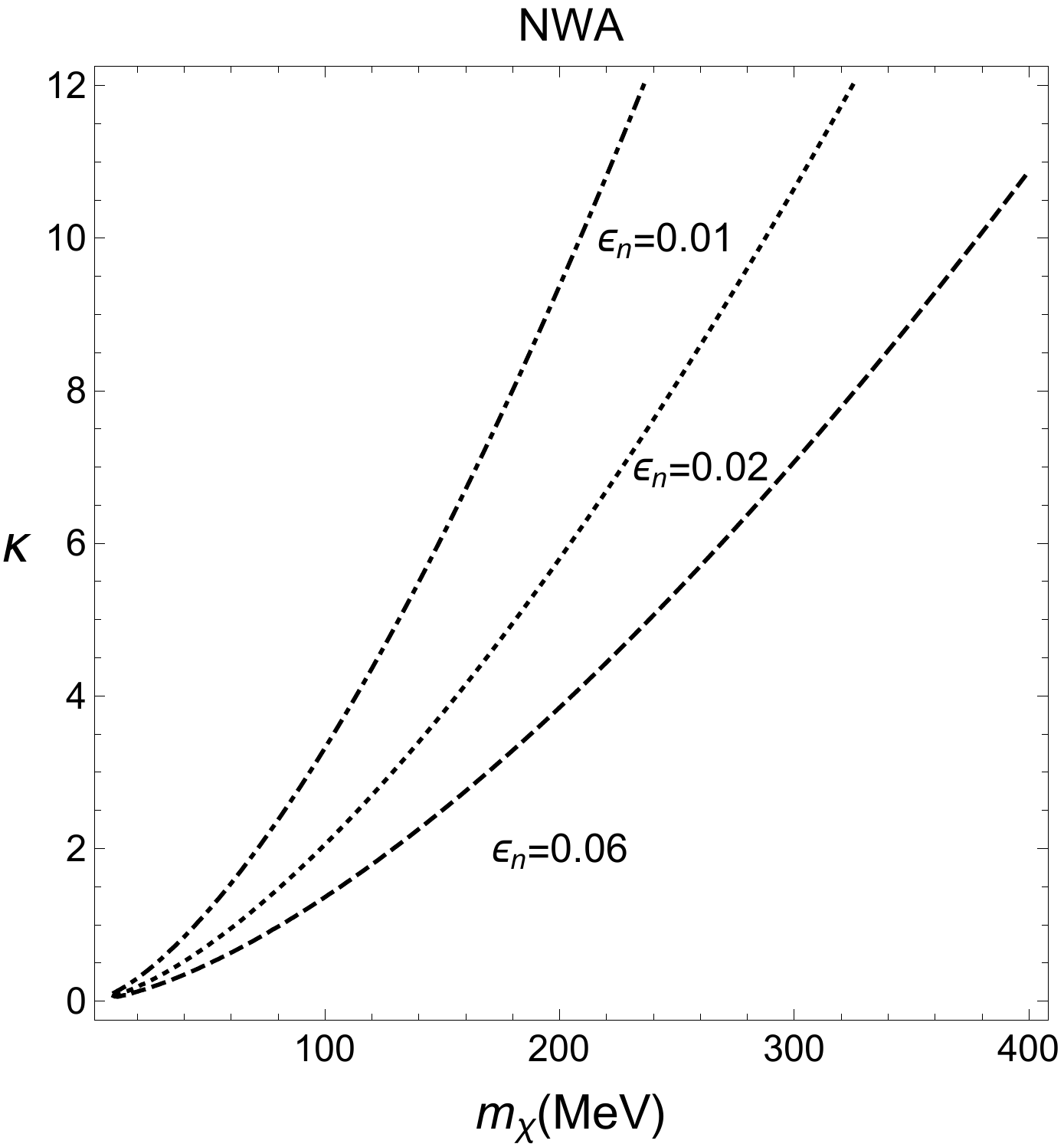}
   \end{center}
  \caption{Parameter space for DM cubic coupling vs mass, satisfying the relic density for $Z_n$ models. The results are shown in the narrow width approximation. $\kappa$ for the $Z_3$ model can be replaced by $\lambda_3$ for the $Z_5$ model.  $\epsilon_n=0.01, 0.02, 0.06$ are chosen from top to bottom lines. }
  \label{Zn}
\end{figure}

Then, since the $3\rightarrow 2$ processes are $s$-wave in all the cases above, using the result in eq.~(\ref{resonance}), we obtain the thermal average as
\be
\langle\sigma v^2\rangle_{Z_n}=\frac{3}{4} C_n \pi x^3 \, G_0(z_n;x),
\ee
with $z_n\equiv \epsilon_n+ i \gamma_n$. 
In the narrow width approximation, the above result becomes
 \be
 \langle\sigma v^2\rangle_{Z_n}\approx \frac{27}{16} C_n \pi\epsilon^2_n \, x^3\,e^{-\frac{3}{2} x \epsilon_n}\,\theta(\epsilon_n). 
 \ee
In Fig.~\ref{swave}, we show the relic density $\Omega h^2$ as a function of $\epsilon_n$ ($C_n$) in the left (right) panel for a fixed $C_n$ ($\epsilon_n$). These results are for the resonance cases with $s$-wave annihilation, which are applicable to $Z_n$ models. DM mass is chosen to $100(200)\,{\rm MeV}$ in the upper (lower) panel. Moreover, in Fig.~\ref{Zn}, we show the parameter space for DM cubic coupling and mass satisfying the relic density measured by Planck, depending on the value of $\epsilon_R=0.01, 0.02, 0.06$ from top to bottom. The DM cubic coupling is given by $\kappa$ for the $Z_3$ model and $\lambda_3$ for the $Z_5$ model. Here, the narrow width approximation is assumed. 
As a consequence, we find that the required value of $\kappa$ for the relic density varies by a factor of $3-5$, depending on $\epsilon_n$. 
We note that we kept only the resonant channels in $Z_n$ models to show the dependence on the resonance pole but extra non-resonant channels to the same $3\rightarrow 2$ process can allow for a smaller $\kappa$ coupling  \cite{z3simp,z5simp,fdm}. Furthermore, other couplings such as $\lambda_\chi$ make the model consistent with the bound on the self-scattering cross section of dark matter \cite{z3simp,z5simp,fdm}.

\subsection{Dark mesons}

We consider non-abelian gauge symmetries with flavor groups in the dark sector,  such as $SU(N_c)$ gauge symmetry and $SU(N_f)\times SU(N_f)/SU(N_f)$ coset space for flavor group. The Wess-Zumino-Witten(WZW) terms \cite{WZW1,WZW2} are responsible for $3\rightarrow 2$ processes for dark mesons \cite{Hochberg2,Hochberg3}. 
When dark quarks are charged under a dark local $U(1)$, the dark gauge boson $Z'$ has vector-like couplings to dark quarks, resulting in dark meson couplings such as $Z'-\pi_i-\pi_j-\pi_k$ and $Z'-\pi_i-\pi_j$ \cite{WZW2,simp2}. In this case, a gauge kinetic mixing between dark photon and SM hypercharge gauge boson allows for dark matter to be in kinetic equilibrium until freeze-out, and the extra $2\rightarrow 2$ (semi-)annihilation channels, $\pi\pi\rightarrow Z' Z'(\pi)$, is kinematically forbidden\footnote{We note that the forbidden channels can be still important for determining the relic density if $m_\pi<m_{Z'}\lesssim 2(\frac{3}{2})m_\pi$ \cite{fdm}. } for $m_{Z'}>m_\chi$.  
Furthermore, the $3\rightarrow 2$ process for dark mesons can have a resonance at $m_{Z'}=3m_\pi$. 

The effective Lagrangian for dark mesons including WZW terms is the following,
\bea
{\cal L}_{\pi}&=& \frac{1}{4} {\rm Tr}\Big(D_\mu\pi (D^\mu \pi)^\dagger \Big)+\frac{2N_c}{15\pi^2 F^5}\,\epsilon^{\mu\nu\alpha\beta} {\rm Tr}\Big(\pi\partial_\mu\pi \partial_\nu\pi \partial_\alpha\pi\partial_\beta \pi \Big)  \nonumber \\
&&+\frac{ig_D N_c}{3\pi^2 F^3}\,\epsilon^{\mu\nu\alpha\beta} Z'_\mu {\rm Tr}\Big(Q_D \partial_\nu \pi\partial_\alpha\pi \partial_\beta \pi \Big) +\cdots
\eea
where $F$ is the decay constant of dark mesons, $\pi\equiv 2T^a \pi^a$ with $T^a$ satisfying $[T^a,T^b]=if_{abc} T^c$ and belonging to $SU(N_f)\times SU(N_f)/SU(N_f)$ (e.g. $\lambda^a=2T^a$ being Gell-Mann matrices for $N_f=3$), and the covariant derivative for dark mesons is given by $D_\mu\pi=\partial_\mu\pi +ig_D Z'_\mu [Q_D,\pi]$. Here, $Q_D$ is the dark charge operator which is chosen to be ${\rm Tr}\,Q_D\neq 0$ and $Q^2_D=1$ for the absence of chiral anomalies \cite{simp2,Hochberg3}. 
For the $3\rightarrow 2$ dominance, we need to suppress $\pi \pi\rightarrow  Z'Z'(\pi)$, requiring $m_{Z'}\gtrsim 2(\frac{3}{2})m_\pi$. 

First, the WZW terms for dark mesons lead to the $d$-wave suppressed $3\rightarrow 2$ processes for dark mesons and the corresponding annihilation cross section takes the following form in the velocity expansion,  
\bea
(\sigma v^2)_{WZW}= C_{WZW} \Big(\frac{1}{4}(v^2_1+v^2_2+v^2_3)^2-\frac{1}{2}(v^4_1+v^4_2+v^4_3) \Big).
\eea
Thus, as there is no resonance, we can make use of eq.~(\ref{dwave}) to get the thermal average as
\bea
\langle\sigma v^2\rangle_{WZW}=2 C_{WZW} x^{-2},
\eea
where $C_{WZW}$ depends on group factors.  The result agrees with Ref.~\cite{Hochberg2}.

The gauged WZW terms for dark mesons lead to additional $3\rightarrow 2$ processes for dark mesons with a resonance. After the dark photon is integrated out, the resulting effective interaction is
\bea
{\cal L}'_\pi=\frac{16 g^2_D N_c}{3\pi^2 m^2_{Z'} F^3}\, {\rm Tr}\Big(Q_D[T^a,T^b] \Big)\,{\rm Tr}\Big(Q_D T^c T^d T^e \Big) \epsilon^{\mu\nu\alpha\beta} \pi^a \partial_\mu \pi^b \partial_\nu \pi^c \partial_\alpha \pi^d \partial_\beta \pi^e.
\eea
For the resonance case, we only have to replace $1/m^2_{Z'}$ by $-1/(s-m^2_{Z'})$ where $s$ is the center of mass energy for $3\rightarrow 2$ processes.
As the gauged WZW terms lead to the effective 5-point interactions of the same form as the one of the ungauged WZW terms, the corresponding $3\rightarrow 2$ annihilation cross section is given by
\bea
(\sigma v^2)_{gWZW}= C_{gWZW} \Big(\frac{1}{4}(v^2_1+v^2_2+v^2_3)^2-\frac{1}{2}(v^4_1+v^4_2+v^4_3) \Big)\,\frac{\gamma_{Z'}}{(\epsilon_{Z'}-\frac{2}{3}\eta)^2+\gamma^2_{Z'}},
\eea
where $\epsilon_{Z'}=\frac{m^2_{Z'}-9m^2_\pi}{9m^2_\pi}$ and $\gamma_{Z'}=\frac{m_{Z'} \Gamma_{Z'}}{9m^2_\pi}$ and $C_{gWZW}$ depends on group factors as well as the dark charge operator $Q_D$. 
Here, the decay rate of the dark photon is approximated by the two-body decay to be
\bea
\Gamma_{Z'}=\frac{g^2_D}{48\pi} {\rm Tr}(\,Q^2_\pi)\, m_{Z'} \Big(1-\frac{4m^2_\pi}{m^2_{Z'}}\Big)^{3/2}.
\eea
Then, in the narrow width approximation, using the result in eq.~(\ref{nwa}) and doing an explicit integration for the thermal average of the terms with $v^2_i v^2_j$, $i\neq j$, we get the thermal average of the additional $3\rightarrow 2$ annihilation cross section as
\bea
\langle\sigma v^2\rangle_{gWZW}= \frac{729}{32} C_{gWZW} \pi  \epsilon^4_{Z'} \,x^3\, e^{-\frac{3}{2}\epsilon_{Z'}}\,\theta(\epsilon_{Z'}). 
\eea
In this case, the resulting averaged cross section has a higher power dependence on $\epsilon_R$ near resonance, due to the overall $d$-wave suppression of the $3\rightarrow 2$ annihilation cross section.

\section{Generalizations}

In this section, we generalize our previous discussion on the thermal average to the cases with non-degenerate masses in the initial states or the $n+2\rightarrow 2$ annihilation processes.

\subsection{$3\rightarrow 2$ co-annihilations}

The results on thermal average can be generalized to the case with non-degenerate masses in the initial states of the $3\rightarrow 2$ process \cite{coann}, namely, the co-annihilation between multiple components of dark matter. In this case, we consider the momenta $p_i(i=1,2,3)$ instead of velocities $v_i(i=1,2,3$) in the integration and the velocity expansion of the $3\rightarrow 2$ annihilation cross section.

For simplicity, we take the $3\rightarrow 2$ annihilation cross section as a function of the total kinetic energy, namely, $K=\frac{p^2_1}{2m_1}+\frac{p^2_2}{2m_2}+\frac{p^2_3}{2m_3}$, in the non-relativistic limit.  
Then, the thermal average for the case with non-degenerate masses can be simply given by the one for the case with degenerate masses where $m_{\rm DM}$ is replaced by $(m_1+m_2+m_3)/3$ in  eqs.~(\ref{etan}) or (\ref{resonance}), depending on whether the process is non-resonant or resonant.
This result is particularly useful for the $s$-wave $3\rightarrow 2$ process with non-degenerate masses.
But, if the $3\rightarrow 2$ co-annihilation process is velocity-suppressed, one needs to take care of the thermal average of all the individual velocity terms, that are not necessarily $SO(9)$ invariant due to mass differences.

\subsection{Higher-order DM annihilations}

We can generalize our previous discussion to the thermal average for $n+2\rightarrow 2$ annihilation processes \cite {Hochberg1,higherorder} with initial particles having the same masses. We denote the corresponding annihilation cross section by $(\sigma v^{n+1})$ and the corresponding thermal average is given by
\bea
\langle \sigma v^{n+1}\rangle= \frac{\int d^3 v_1 \cdots d^3 v_{n+2}\, \delta^3({\vec v}_1+\cdots+{\vec v}_{n+2})(\sigma v^{n+1})\, e^{-\frac{1}{2}x(v^2_1+\cdots+v^2_{n+2})}}{\int d^3 v_1\cdots d^3 v_{n+2}\, \delta^3({\vec v}_1+\cdots+{\vec v}_{n+2}) \,e^{-\frac{1}{2}x(v^2_1+\cdots+v^2_{n+2})}}.
\eea
Then, in the case of the $SO(3(n+2))$ invariant velocity expansion, namely, $(\sigma v^{n+1})=\sum_{l=0}^\infty \frac{a_l}{l!}\,\eta^l$ with $\eta=\frac{1}{2}(v^2_1+\cdots+v^2_{n+2})$, we obtain the thermal average in a simple matter as
\bea
\langle \sigma v^{n+1}\rangle&=& \frac{x^{\frac{3}{2}(n+1)}}{\Gamma(\frac{3}{2}(n+1))}\int^\infty_0 d\eta\, \eta^{\frac{1}{2}(3n+1)+l}\, e^{-x\eta} \nonumber \\
&=& \frac{1}{\Gamma(\frac{3}{2}(n+1))} \sum_{l=0}^\infty \Gamma\Big(\frac{3}{2}(n+1)+l\Big)\frac{a_l}{l!}\,\,x^{-l}.
\eea
Likewise in the case of $3\rightarrow 2$ processes, in most cases, the most important terms appear up to $p$-wave terms that are $SO(3(n+2))$ invariant, so the above result gives rise to a good approximation for the full average of $n+2\rightarrow 2$ processes.

\section{Conclusions}

We have presented general results on the thermal average of $3\rightarrow 2$ annihilation cross sections of dark matter. 
The results can be important to improve the calculation of the dark matter abundances in the case with strong velocity-dependence and resonance poles. We have shown some examples on SIMP dark matter where the obtained results can be applied and have extended our discussion to  the case with the $3\rightarrow 2$ co-annihilation and even higher-order annihilation processes.

\section*{Acknowledgments}

We would like to thank the CERN Theory group for hospitality and participants in the CERN Theory Institute on ``New Physics at Intensity Frontiers'' for discussion during the final stage of the project. 
We would like to thank Myeonghun Park for the discussion on three-body decays. 
The work is supported in part by Basic Science Research Program through the National Research Foundation of Korea (NRF) funded by the Ministry of Education, Science and Technology (NRF-2016R1A2B4008759). The work of SMC is supported in part by TJ Park Science Fellowship of POSCO TJ Park Foundation.
MS is supported by IBS under the project code, IBS-R018-D1.

\end{document}